\documentclass[a4paper,fleqn,usenatbib]{mnras}

\usepackage{mathptmx}

\usepackage[T1]{fontenc}
\usepackage{ae,aecompl}

\usepackage{graphicx}	
\usepackage[caption=false]{subfig}
\usepackage{amsmath}	
\usepackage{amssymb}	
\usepackage[normalem]{ulem}

\usepackage{multirow}

\newcommand{\fielddag}{{\hat \phi}^\dagger}
\newcommand{\field}{{\hat \phi}}
\newcommand{\de}[1]{\partial_{#1 }}
\newcommand{\sumnn}{\sum_{j\in \text{NN}(i)}}
\newcommand{\Lagr}{\mathcal{L}}
\newcommand{\G}{{\small P-GADGET}3 }
\newcommand{\AG}{{\small AX-GADGET} }

\renewcommand{\vec}[1]{\mathbf{#1}}

\title[AX-GADGET: a new code for Fuzzy Dark Matter models]{AX-GADGET: a new code for cosmological simulations of Fuzzy Dark Matter and Axion models}

\author[Matteo Nori et al.]{
Matteo Nori$^{1,2,3}$\thanks{E-mail: matteo.nori3@unibo.it}
and Marco Baldi$^{1,2,3}$\thanks{E-mail: marco.baldi5@unibo.it}
\\
$^{1}$Dipartimento di Fisica e Astronomia, Alma Mater Studiorum - University of Bologna, Via Piero Gobetti 93/2, 40129 Bologna BO, Italy\\
$^{2}$INAF - Osservatorio Astronomico di Bologna, Via Piero Gobetti 93/3, 40129 Bologna BO, Italy\\
$^{3}$INFN - Istituto Nazionale di Fisica Nucleare, Sezione di Bologna, Viale Berti Pichat 6/2, 40127 Bologna BO, Italy
}

\date{Accepted XXX. Received YYY; in original form ZZZ}

\pubyear{2018}

\begin{document}
\label{firstpage}
\pagerange{\pageref{firstpage}--\pageref{lastpage}}
\maketitle

\begin{abstract}
We present a new module of the parallel N-Body code \G for cosmological simulations of light bosonic non-thermal dark matter, often referred as Fuzzy Dark Matter (FDM). The dynamics of the FDM features a highly non-linear Quantum Potential (QP) that suppresses the growth of structures at small scales. Most of the previous attempts of FDM simulations either evolved suppressed initial conditions, completely neglecting the dynamical effects of QP throughout cosmic evolution, or resorted to numerically challenging full-wave solvers. The code provides an interesting alternative, following the FDM evolution without impairing the overall performance. This is done by computing the QP acceleration through the Smoothed Particle Hydrodynamics (SPH) routines, with improved schemes to ensure precise and stable derivatives. As an extension of the \G code, it inherits all the additional physics modules implemented up to date, opening a wide range of possibilities to constrain FDM models and explore its degeneracies with other physical phenomena. Simulations are compared with analytical predictions and results of other codes, validating the QP as a crucial player in structure formation at small scales. 
\end{abstract}




\section{Introduction}

According to the currently accepted standard cosmological scenario -- known as the $\Lambda $CDM model -- about $80\%$ of the matter content of the universe is in the form of Cold and collisionless Dark Matter particles (CDM), whose contribution to the gravitational instability of density perturbations drives the formation of present cosmic structures stemmed from the tiny primordial fluctuations observed in the Cosmic Microwave Background \citep[CMB, see e.g. ][]{Planck15}.

The success of the $\Lambda $CDM model has been significantly supported over the past decades by the development and continuous improvement of numerical techniques, that allowed us to simulate the evolution of cosmic structures in an expanding universe from the well understood linear domain, constrained by CMB observations, down to the highly non-linear regime that characterises the present-day density field at small scales. In this respect, large and sophisticated cosmological N-body simulations -- as well as their hydrodynamical extensions accounting for the complex astrophysical processes, related to the subdominant baryonic matter component -- have undeniably become an essential tool in contemporary astrophysics and cosmology.

Although the existence of dark matter is a solid outcome of a large number of independent observations -- as e.g. the inner dynamics of galaxy clusters \citep[][]{Zwicky_1937,Clowe06}, the rotation curves of spiral galaxies \citep[][]{Rubin_Ford_Thonnard_1980,Bosma_1981,Persic96}, the strong  gravitational lensing of individual massive objects \citep[][]{Koopmans02} as well as the weak gravitational lensing arising from the large-scale matter distribution \citep[][]{Mateo98,Heymans_etal_2013,Planck_2015_Gravitational_Lensing,Hildebrandt_etal_2017}, the angular power spectrum of CMB temperature anisotropies \citep[as observed e.g. from WMAP and Planck][respectively]{wmap7,Planck_2015_XIII}, the clustering of luminous galaxies \citep[see e.g.][]{VIPERS_Om_M,SDSS-III-final}, the large-scale velocity field \citep[][]{Bahcall98} and the abundance of massive clusters \citep[][]{Kashlinsky98} -- and despite most of the proposed alternatives based on ad-hoc modifications of gravity \citep[such as Modified Newtonian Dynamics and its variants, see e.g.][]{Milgrom_1983,Sanders02,Bekenstein04} have been recently ruled out \citep[][]{Chesler17} by the implications of the gravitational wave event GW170817 \citep[][]{GW170817},  the fundamental nature of dark matter is far from being understood and consistent direct or indirect detections of plausible dark matter particle candidates have been eluding our attempts so far.

The lack of a positive result in the decades-long hunt for dark matter particles is now starting to undermine the popularity of the most massive candidates, like Weakly Interactive Massive Particles (WIMPs) arising in supersymmetric extensions of the standard model of particle physics (as e.g. the {\em neutralino}), hence providing motivations to explore alternative scenarios often characterised by lighter particle species \citep[see e.g.][for an excellent review on Dark Matter particle candidates]{Bertone_Hooper_Silk_2005}.

Furthermore, it is still highly debated whether the apparent failures of CDM models at scales $\lesssim 10\ kpc$ -- as given e.g. by the cusp-core problem \citep[][]{Oh11}, the missing satellite problem \citep[][]{Klypin99}, the too-big-to-fail problem \citep[][]{Boylan-Kolchin12} -- may be related to an imperfect baryonic physics implementation in numerical simulations \citep[see e.g.][]{Maccio12,Brooks13}, to an intrinsic diversity of properties related to the formation history and local environment of each individual dark matter halo \citep[][]{Oman_etal_2015}, or ultimately to the fundamental nature of the dark matter particle \citep[see e.g.][]{Spergel00,Rocha13,Kaplinghat00,Medvedev14}.

One intriguing solution to these problems might involve an extremely light non-thermal boson acting as dark matter, whose de-Broglie wavelength arising from its fundamental quantum nature would be relevant at cosmological scales \citep[see e.g. ][]{Marsh10,Hui16}. The lightness and quantum behaviour of such bosonic dark matter particles could simultaneously explain its elusiveness and alleviate tensions at small scales \citep[see e.g. ][]{Marsh15CCP}. This type of dark matter has been generically termed {\em Fuzzy Dark Matter}  \citep[FDM hereafter, see ][]{Hu00} and several particles that fit in this description have been proposed in the literature, the most popular class being Ultra Light Axions \citep[ULAs, ][]{Marsh16}.


With the next generation of cosmological surveys \citep[as e.g. Euclid, LSST, SKA, see][, respectively]{EUCLID,LSST,SKA} starting to take data in the near future, holding the promise to pinpoint with unprecedented precision the parameters involved in the $\Lambda$CDM model and to detect even extremely feeble signals of deviations from the standard cosmology, the urge for more accurate predictions on the expected signatures of alternative scenarios is now a high priority for the community. In this respect, developing numerical tools for cosmological simulations of alternative dark matter candidates represents a necessary step to provide such predictions.

In this paper, we present a modification of the cosmological N-Body and hydrodynamical code \G -- a non-public extension of the public {\small GADGET}2 code \citep[][]{Springel05} -- to simulate the non-linear evolution of FDM scenarios featuring light boson fields as dark matter particles.

In these models the dynamics of FDM particles is influenced -- besides gravity -- by an additional {\em Quantum Potential} that we are able to compute exploiting the Smoothed Particle Hydrodynamic routines \citep[as suggested by e.g. ][]{Mocz15,Marsh15} already implemented in \G for standard hydrodynamical simulations. This method allows us to keep track of the Quantum Potential effects into the non-linear regime of structure formation while keeping under control the overall computational performances, which makes the code more suitable for large cosmological simulations compared to other methods resorting on full quantum wave solvers \citep[such as the grid-based algorithms presented in][]{GAMER}.

Our implementation is flexible enough to easily include models with dark matter self-interaction and allows multiple dark matter species, either fuzzy or not. However, in the present work we focus on the case of a single FDM component accounting for the total dark matter budget.
We discuss and compare the results of our algorithm to analytical solutions and to the recent results of other similar codes, in order to point out the reliability of our algorithm, its overall performance, as well as the predicted effect of the Quantum Potential on the statistical and structural properties of cosmic structures.

\ \\

The paper is organized as follows: in section~\ref{sec:theory} we briefly describe the Axion Dark Matter models under consideration, providing all the basic equations that enter our numerical implementation, and reviewing the physical dynamics of light bosonic fields (\ref{sec:maths}). We then present the implementation of such equations into the code (\ref{sec:code}). In section~\ref{sec:cv} we analyse the results of a series of tests for idealised setups having known analytical solutions as well as for cosmological simulations (\ref{sec:sims}), and we present the overall performance of our code (\ref{sec:perf}). Finally, in section~\ref{sec:conclusions} we draw our conclusions.

\section{Theory and Algorithm}
\label{sec:theory}

In this section, we recall the dynamics of a light bosonic field and its formulation in terms of an effective fluid, by introducing the corresponding hydrodynamic equations, and we present the implementation strategy we used in the code to solve for such equations, based on the Smoothed Particle Hydrodynamics (SPH) approach.

\subsection{Dynamics}
\label{sec:maths} 

Let $\field$ be a bosonic field evolving accordingly to the Gross-Pitaevskii-Poisson equation \citep[][]{Gross61,Pitaevskii61}
\begin{equation}
\label{eq:GPP}
i \hbar \ \de{t} \field = - \frac {\hbar^2} {m_{\chi}^2} \nabla^2 \field + m_\chi \Phi \field + \lambda \left( \fielddag \field \right) \field
\end{equation}
where $\Phi$ is the Newtonian gravitational potential, $\lambda$ and $m_{\chi}$ represent the self-interaction coupling constant and typical mass of the field, respectively.

In order to describe the dynamics of such field in terms of fluid equations, we use the Madelung form \citep[][]{Madelung27}
\begin{equation}
\field = \sqrt {\frac \rho m_{\chi}} e^{i \frac \theta \hbar}
\end{equation}
where $\rho$ is the fluid density and $\theta$ is related to the fluid velocity as $\vec v = \vec \nabla \theta / m_{\chi}$. Extending this approach to the case of an expanding universe -- in a comoving frame with $a$ and $H=\dot a / a$ being the usual cosmological scale factor and Hubble parameter, respectively -- we recover the well known Madelung equations, consisting in the continuity equation
\begin{equation}
\label{eq:continuity}
\dot \rho + 3 H \rho + \frac 1 a \vec \nabla \cdot \left( \rho \vec v \right) =0
\end{equation}
and a modified Navier-Stokes equation
\begin{equation}
\label{eq:NS}
\dot {\vec v} + \frac 1 a \left( \vec v \cdot \vec \nabla \right) \vec v =  - \frac {\vec \nabla \Phi} a  + \frac {\vec \nabla P} {a \rho}+ \frac {\vec \nabla Q} {a^3} 
\end{equation}
where three distinct sources of particle acceleration appear: the gravitational potential $\Phi$, a pressure-like term $P$ accounting for the self-interaction of the field, and an additional potential $Q$.

The gravitational potential $\Phi$ satisfies the usual Poisson equation 
\begin{equation}
\label{eq:poisson}
\nabla^2 \Phi = 4 \pi G a^2 \rho_b \ \delta
\end{equation}
where $\delta=(\rho - \rho_b)/\rho_b$ is the density contrast with respect to the background field density $\rho_b$ \citep{Peebles80}.

The self-interaction term $P$ can be regarded as a pressure and, in principle, can be generalised with a parametric equation of state $P=P(\rho)$ to take into account self-interactions of different forms. In the case of quartic self-interaction, as in Eq.~\ref{eq:GPP}, it reads $P=\lambda \rho^2 / 2 m_\chi$.

The potential $Q$ has the form of
\begin{equation}
\label{eq:QP}
Q = \frac {\hbar^2}{2m_{\chi}^2} \frac{\nabla^2 \sqrt{\rho}}{\sqrt{\rho}}  = \frac {\hbar^2}{2m_{\chi}^2} \left( \frac {\nabla^2 \rho} {2 \rho} - \frac {| \vec \nabla \rho|^2}{4 \rho^2} \right)
\end{equation}
and we acknowledge its use in the literature since the 50s as \textit{Quantum Potential} (QP) \citep{Bohm52}.
In recent applications in cosmology, it has been expressed sometimes as a pressure tensor 
\begin{equation}
\nabla Q = \frac 1 \rho \nabla P_Q %
= \frac {\hbar^2}{2m_{\chi}^2} \ \frac 1 \rho \nabla \left( \frac \rho 4 \nabla \otimes \nabla \ln{\rho} \right)
\end{equation}
thus addressed as \textit{Quantum Pressure} \citep[see e.g. ][]{Mocz15}. In this work we prefer the former \textit{potential} terminology, given its uninvolvement with classical thermal interactions. We find necessary, indeed, to stress that this potential has neither links with temperature nor any classical thermodynamics origin. Its mathematical form and physical behaviour are related  to a self-organizing process and are connected to basic principles of quantum information and occupation of states \citep[][]{Boehmer07}.

The set of Eq.~\ref{eq:continuity}-\ref{eq:poisson} has a stable solution for $\delta = \Phi = |\vec v| = 0$ which can be perturbed assuming $\delta, \Phi, |\vec v| \ll 1$ in order to end up with a new set of linearised equations. The resulting equations can be combined in the density contrast time evolution reading 
\begin{equation}
\label{eq:pert}
\ddot \delta + 2 H \dot \delta + \left(  \frac{\hbar^2 k^4}{4 m_{\chi}^2 a^4} + \frac{c_s^2 k^2}{a^2} - \frac{4 \pi G \rho_b } {a^3} \right) \delta = 0
\end{equation}
where $\delta(\vec x,t)$ has been decomposed in Fourier modes $\delta_k e^{i \vec k \cdot \vec x}$ and $c_s^2=\de{\rho} P(\rho)|_{\rho_b}$ is the sound speed of the fluid \citep{Chavanis12}.

Our implementation (see below) is able to simulate models with any given self-interaction encoded in $P(\rho)$ as a parametrised input. However, in this paper we restrict our focus only on the effects of the QP and leave the exploration of self-interacting models for future work. We therefore consider $\lambda=0$, thus $P=c_s=0$, hereafter.

A perturbed stable solution of Eq.~\ref{eq:pert}, is given by a density contrast with mode
\begin{equation}
\label{eq:kq}
k_Q (a)= \left( \frac{16\pi G \rho_b a^3 m_{\chi}^2}{\hbar^2} \right)^{1/4} a^{1/4} 
\end{equation}
and corresponding wavelength $\lambda_Q = 2 \pi / k_Q$, representing a quantum version of the Jeans wavenumber and Jean length, respectively.

Given the cosmological scale factors $a(t)$ at each time $t$, the wavelengths $\lambda_Q$ represent the scale at which gravity is perfectly balanced by the QP, dividing a region of gravitational collapsing instability ($\lambda>\lambda_Q$) from a region of expansion (for $\lambda<\lambda_Q$) due to the net repulsive effect of the QP \citep[][]{Chavanis12,Woo09}.

The general solution of Eq.~\ref{eq:pert} can be expressed as a linear combination of a growing mode $D_+(k,a)$ and a decaying mode $D_-(k,a)$ that have the form
\begin{equation}
\begin{aligned}
\label{eq:growth}
D_+(x) = \left[ \left( 3 - x^2 \right) \cos{x} + 3\ x \sin{x}\right] / x^2 \\
D_-(x) = \left[ \left( 3 - x^2 \right) \sin{x} - 3\ x \cos{x}\right] / x^2
\end{aligned}
\end{equation}
where we defined the parameter $x(k,a)=\sqrt{6}\ k^2 / k_Q^2(a)$. The linear solution, therefore, predicts a suppression of structures in the density field on small scales -- i.e. for $k \gg k _{Q}$ -- in which both the growing and decaying mode oscillate in time, effectively halting density perturbation evolution. At large scales -- i.e. for $k \ll k _{Q}$ -- the standard linear evolution $D_ + \propto a$ and $D_- \propto a^{-2/3}$ of CDM is recovered thus allowing density perturbation growth \citep[][]{Hu00}.

The presence of an oscillating regime gives rise to a cutoff of the small-scale density power spectrum similarly to what happens for Warm Dark Matter particle candidates \citep[][]{Bode00}. However, the mechanisms generating such effects in the two cases have a completely different origin, resulting in a different shape of the respective transfer functions.

Since the quantum Jeans wavenumber $k_Q(a) \propto a^{1/4}$ increases with time as from Eq.~\ref{eq:kq}, we expect oscillating modes to eventually start growing, each at a different redshift, as they are passed by the quantum Jeans scale. While, in the linear approximation, the fastest possible growth for density perturbation of Eq.~\ref{eq:growth} is the one that characterizes the largest scales $D_+ \propto a$ -- making it impossible for the intermediate suppressed scales to catch up with the largest ones -- in the non-linear regime we expect a faster growth for the intermediate and small scales at low redshift, allowing such restoring effect.

Therefore, it is necessary to resort on numerical techniques to investigate the detailed integrated effects of these scenarios, and to develop suitable codes to perform N-Body hydrodynamical simulations that could follow their evolution deep into the fully non-linear regime.

\subsection{Implementation}
\label{sec:code}

The \AG code we developed relies on SPH techniques -- already partly implemented in \G -- to solve for the QP, computed for each fluid particle through local summation algorithms using Eq.~\ref{eq:QP}, and adopt it as additional source of acceleration in the Navier-Stokes equation, as suggested by \citet{Mocz15,Marsh15}.
To this end, we have equipped dark matter particles in \AG with an additional data structure to store the necessary quantities that are relevant for the fluid representation of the FDM and similar to the one already in place for gas particles. As for the native SPH implementation of \G, the exchange between CPUs of such additional layer of data for local particles is optimized to guarantee high memory efficiency in the domain decomposition.

SPH provides us with a numerical strategy to approximate continuous fields with sums over neighbouring particles. In such approximations, the deviation of numerical results from the exact solution depends on several factors, related in particular to the intrinsic limitations of a Lagrangian particle description of fluids, where shocks and strong interface interactions tend to be smoothed out or underestimated. However, the flexibility of the method allows to rearrange the specific form of the SPH machinery such that equivalent analytical problems can be implemented into flavours of the basic algorithms with different levels of numerical accuracy, as in detail below.

The basic concept behind the SPH approach resides in expressing the value of a given observable $O$ at the position of particle $i$ as the sums of its value over $\text{NN}(i)$ neighbouring particles
\begin{equation}
O_i = \sumnn m_j \frac {O_j} {\rho_j} W_{ij}
\end{equation}
weighted on mass $m_j$, density $\rho_j$ and a window function $W_{ij}$ -- sometimes referred as kernel function --, which can take many possible functional forms. Consequently, the derivative of the observable can be applied to the window function as
\begin{equation}
\label{eq:1SPH}
\vec \nabla O_i = \sumnn m_j \frac {O_j} {\rho_j} \vec \nabla W_{ij} 
\end{equation}
which, however, does not guarantee that $\vec \nabla O$ vanishes for constant values of $O$. 

Therefore, to ensure that the overall derivative vanishes in the case of $O_j \rightarrow O_i \ \forall j \in \text{NN}(i)$, we consider a differentiable function $\Theta$ such that 
\begin{equation}
\vec \nabla O = \frac 1 \Theta \left[ \vec \nabla \left( \Theta\ O \right) - O \ \vec \nabla\ \Theta \right]
\end{equation}
that is translated in the SPH algorithm as
\begin{equation}
\label{eq:1SPHcorrect}
\vec \nabla O_i = \sumnn m_j \frac {O_j - O_i} {\rho_j}  \frac {\Theta_j}{\Theta_i} \vec \nabla W_{ij}
\end{equation}
where the difference of Eq.~\ref{eq:1SPHcorrect} with respect to Eq.~\ref{eq:1SPH} fulfills the condition of null derivative in the case of a constant field, regardless the form of the function $\Theta$. In the literature, $\Theta=1$, $\Theta=\rho$ and $\Theta=\sqrt{\rho}$ are the most common choices \citep[see e.g][]{Monaghan05}. For the different forms of $\Theta$, the derivative of the density field then takes the form
\begin{equation}
\vec \nabla \rho_i = \sumnn m_j  \vec \nabla W_{ij} \left( \rho_j - \rho_i \right) \begin{cases}
1 / \rho_i & \text{for $\Theta=1$} \\
1 / \rho_j & \text{for $\Theta=\rho$} \\
1 / \sqrt{\rho_i \rho_j} & \text{for $\Theta=\sqrt{\rho}$}
\end{cases}
\end{equation}

We noticed that $\Theta=\sqrt{\rho}$ performs better with respect to the other two possibilities in cosmological simulations, where it is not uncommon to find fluid particles with neighbours with quite different density, for example in collapsing regions. Indeed, $\Theta=1$ or $\Theta=\rho$ may lead to high and non-symmetrical $i \rightleftharpoons j$ correction factors that are more sensitive to noise and less likely to disappear in the sum, while $\Theta=\sqrt{\rho}$ translates in a more stable algorithm, which makes it our preferred choice. Nevertheless, the $\Theta$ functional forms can be selected upon compilation in our code, so that the final choice is left open.
\ \\

Even if the first derivative in the form of Eq.~\ref{eq:1SPH} is quite common in the literature, there is no general consensus about the SPH form of the Laplacian.

A straightforward and standard approach consists in applying directly the operator to the window function
\begin{equation}
\label{eq:LSPHnocorrection}
\nabla^2 O_i = \sumnn m_j \frac {O_j} {\rho_j} \nabla^2 W_{ij} ,
\end{equation}
but, in general, such simple implementation leads to unstable results that are very sensitive to irregularities in the particle distribution, mainly due to the steepness of the second derivative of the kernel function.

For the cases under investigation in the present work, results obtained with this implementation for the computation of the QP led us to unsatisfying results. In fact, without resorting to any kind of functional correction, the contribution of the Laplacian term $\nabla^2 \rho / \rho$ to the QP in Eq.~\ref{eq:QP} was negligible compared to $|\vec \nabla \log{\rho}|^2$ and thus positive contributions to the QP were underestimated.

A solution to this problem can be derived employing the same approach described above for the first derivative to improve Eq.~\ref{eq:LSPHnocorrection}, expanding through product derivatives the Laplacian operator as 
\begin{equation}
\label{eq:dprod}
\nabla^2 O = \frac 1 \Theta \left[ \nabla^2 (\Theta\ O) - O \ \nabla^2 \Theta - 2\ \vec \nabla O \cdot \vec \nabla \Theta \right]
\end{equation}
thereby obtaining the correction
\begin{equation}
\label{eq:LSPH}
\nabla^2 O_i =  \sumnn m_j \frac {O_j - O_i} {\rho_j}  \frac {\Theta_j}{\Theta_i} \nabla^2 W_{ij} - \frac {2}{\Theta_i} \vec \nabla O_i \cdot \vec \nabla \Theta_i
\end{equation}
where $\Theta$ was chosen coherently with the density correction \citep[for a comparison between different algorithms see][]{Colin06}. 

A widely used alternative to approximate the Laplacian relies on the symmetry properties of the window function and was first described in \citet[][]{Brookshaw85}
\begin{equation}
\label{eq:BSPHnocorrection}
\nabla^2 O_i = - 2 \sumnn m_j \frac { O_j - O_i  } {\rho_j} \ \frac { \vec r_{ij} \cdot \vec \nabla W_{ij} } {|r_{ij}|^2}
\end{equation}
being one of the most common strategy due to its computational efficiency deriving from the direct dependence on the first derivative of the kernel \citep[see e.g.][]{Cleary99,Jubelgas04,Szewc12}. It is easy to see that, using Eq.~\ref{eq:dprod}, an improved version of such scheme reads
\begin{equation}
\label{eq:BSPH}
\nabla^2 O_i = - 2 \sumnn m_j \frac { O_j - O_i  } {\rho_j} \frac {\Theta_j}{\Theta_i} \ \frac { \vec r_{ij} \cdot \vec \nabla W_{ij} } {|r_{ij}|^2} - \frac {2}{\Theta_i} \vec \nabla O_i \cdot \vec \nabla \Theta_i
\end{equation}
that, apart from the window function dependence, shares a similar structure and the same computational cost with Eq.~\ref{eq:LSPH}.

The numerical errors of each specific algorithm are linked to the numerical instabilities that may arise from the consecutive derivatives of the window function and the compensation of derivative residuals. 
To determine which scheme was the most suitable for our work, we investigated different representations of the density gradient and Laplacian.

\begin{figure}
\includegraphics[width=\columnwidth,trim={0.4cm 0.3cm 1.5cm 1.5cm},clip]{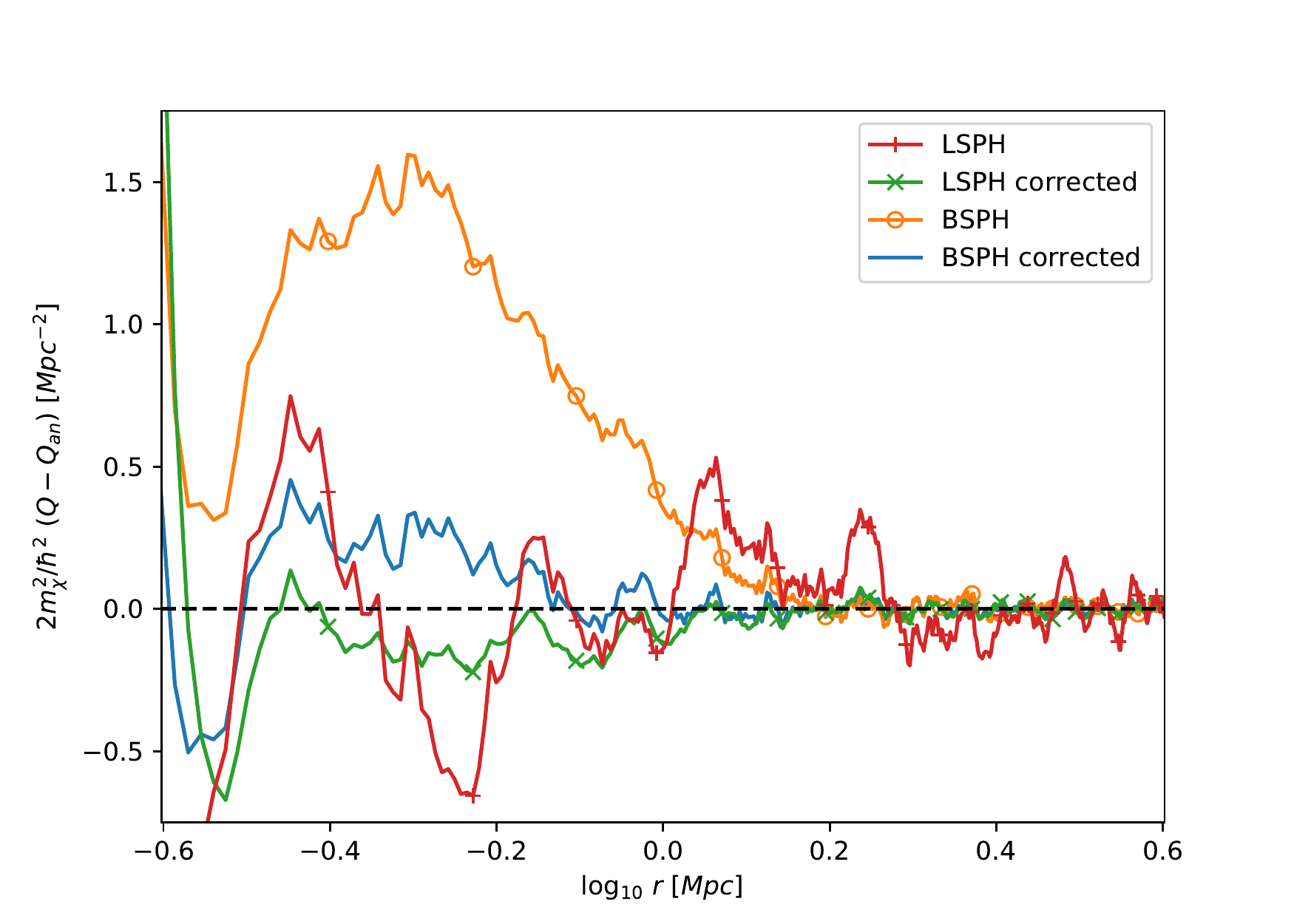}
\caption{Quantum Potential residuals obtained with different algorithms for a 3D Gaussian density distribution. The results  displayed are derived with the Laplacian scheme (LSPH) of Eq.~\ref{eq:LSPHnocorrection}--\ref{eq:LSPH} and with the {\it \`{a} la} Brookshaw (BSPH) scheme of Eq.~\ref{eq:BSPHnocorrection}--\ref{eq:BSPH}, both presented with and without derivative corrections.}
\label{fig:SPH_TESTS}
\end{figure}

In Fig.~\ref{fig:SPH_TESTS} are shown the QP residuals with respect to the analytical results obtained for a 3D Gaussian density distribution -- described in detail in section~\ref{sec:gauss} -- with different schemes. Both the standard and corrected version of the Laplacian implementation ({\it LSPH} and in the figure) of Eq.~\ref{eq:LSPHnocorrection}--\ref{eq:LSPH} and of the {\it \`{a} la} Brookshaw implementation ({\it BSPH} in the figure) of Eq.~\ref{eq:BSPHnocorrection}--\ref{eq:BSPH} are shown. It is undeniable the improvement provided by the correction of the derivatives. Once corrected, the two approaches produce a similar result -- due to their analogous structure -- that follow well the analytic result, in particular the Laplacian algorithm approaches it from below while the other tends to overestimate it.

We hereafter adopt as our preferred algorithm the corrected Laplacian version of Eq.~\ref{eq:LSPH} as it statistically performs better and, as for the $\Theta$ function, the other scheme can be selected upon compilation of the code.
We further discuss and motivate our choices, providing some comparisons and analytical tests, in section~\ref{sec:cv}.
\ \\

The window function used in the code is the cubic B-Spline routinely employed in \G SPH simulations:
\begin{equation}
W(r,h) = \frac 8 {\pi h^3} \begin{cases}
1 - 6 \left(r/h \right)^2 + 6 \left( r/h \right)^3 & \text{if $0<r/h\leq1/2$} \\
2\ \left( 1 - r/h \right)^3 & \text{if $1/2<r/h<1$} \\
0 & \text{otherwise}
\end{cases}
\end{equation}
with $r$ and $h$ being the distance between particles and the smoothing length, respectively. We denote $W_{ij} = W(r_{ij}=|\vec{r_j}-\vec{r_i}|,h_i)$. 
Other higher order functionals are implemented in \G as the quintic B-Splines or the $C4$ and $C6$ Wendland functions \citep[][]{Wendland95} which, however, require a higher number of neighbours to be taken into account \citep[see e.g. for comparison between window functions][]{Dehnen12}. We find no critical variation in accuracy for the calculation of the QP and its derivative when different window functions are used.

Following the standard SPH approach, the value of the single particle smoothing length $h_i$ is varied at each timestep to satisfy the condition
\begin{equation}
\label{eq:NN}
\frac 4 3 \pi h_i^3 \rho_i= \sumnn m_j = M
\end{equation}
such that its corresponding sphere encloses enough neighbours $\text{NN($i$)}$ to match a given amount of mass $M$. With $h$ let free to vary, the condition above can be enforced through a Lagrangian multiplier and $h$-derivatives enter the equations, as described in \citet[][]{Springel01}. For a general presentation we describe such contribution to the QP as factors $f$ (the detailed derivation can be found in Appendix~\ref{sec:dhsml}). 

Given that the quantum acceleration $\vec \nabla Q$ is proportional to a third order derivative of the density field, it is impossible to build an iterative SPH algorithm with less than three cycles over all particles. Therefore, our implementation can be schematically summarized by three cycles of computation.
The first one for the density:
\begin{gather}
\rho_i = \sumnn m_j W_{ij},
\end{gather}
the second one for its gradient and Laplacian:
\begin{gather}
\vec \nabla \rho_i = \sumnn m_j \vec \nabla W_{ij} \frac {\rho_j  - \rho_i} {\sqrt{\rho_i \rho_j}} \\
\nabla^2 \rho_i =  \sumnn m_j \nabla^2 W_{ij}  \frac {\rho_j  - \rho_i} {\sqrt{\rho_i \rho_j}} - \frac {|\vec \nabla \rho_i|^2} {\rho_i} ,
\end{gather}
and the third one to build, using Eq.~\ref{eq:QP}, the QP contribution to acceleration:
\begin{equation}
\vec \nabla Q_i =  \frac {\hbar^2}{2m_{\chi}^2} \sumnn \frac {m_j} {f_j \rho_j} \vec \nabla W_{ij} 
\left( \frac {\nabla^2 \rho_j} {2 \rho_j} - \frac {| \vec \nabla \rho_j|^2}{4 \rho_j^2} \right).
\end{equation}

It would be perfectly legitimate, however, to choose different ways to break down the derivatives into lower order sums on neighbouring particles following the SPH prescriptions.
Regardless of the specific formulation chosen, the algorithm should always be tested against known analytical solutions, in order not to trade better performances off for worse results. In the next Section we present a series of basic tests of our FDM implementation showing that the original SPH formulation results in a very poor accuracy of the numerical solution for the QP, while our improved strategy provides much better results.

\section{Code Validation}
\label{sec:cv}

In this section, we test the accuracy of the algorithm described in the previous section by comparing, for some particular density distributions, the solution of the QP obtained from our implementation to the analytic one. We also compare our improved SPH scheme for spatial derivatives with the results obtained through the standard SPH implementation of \G.

\begin{figure*}
\includegraphics[width=0.246\textwidth,trim={4.6cm 1.3cm 5.1cm 3.8cm},clip]{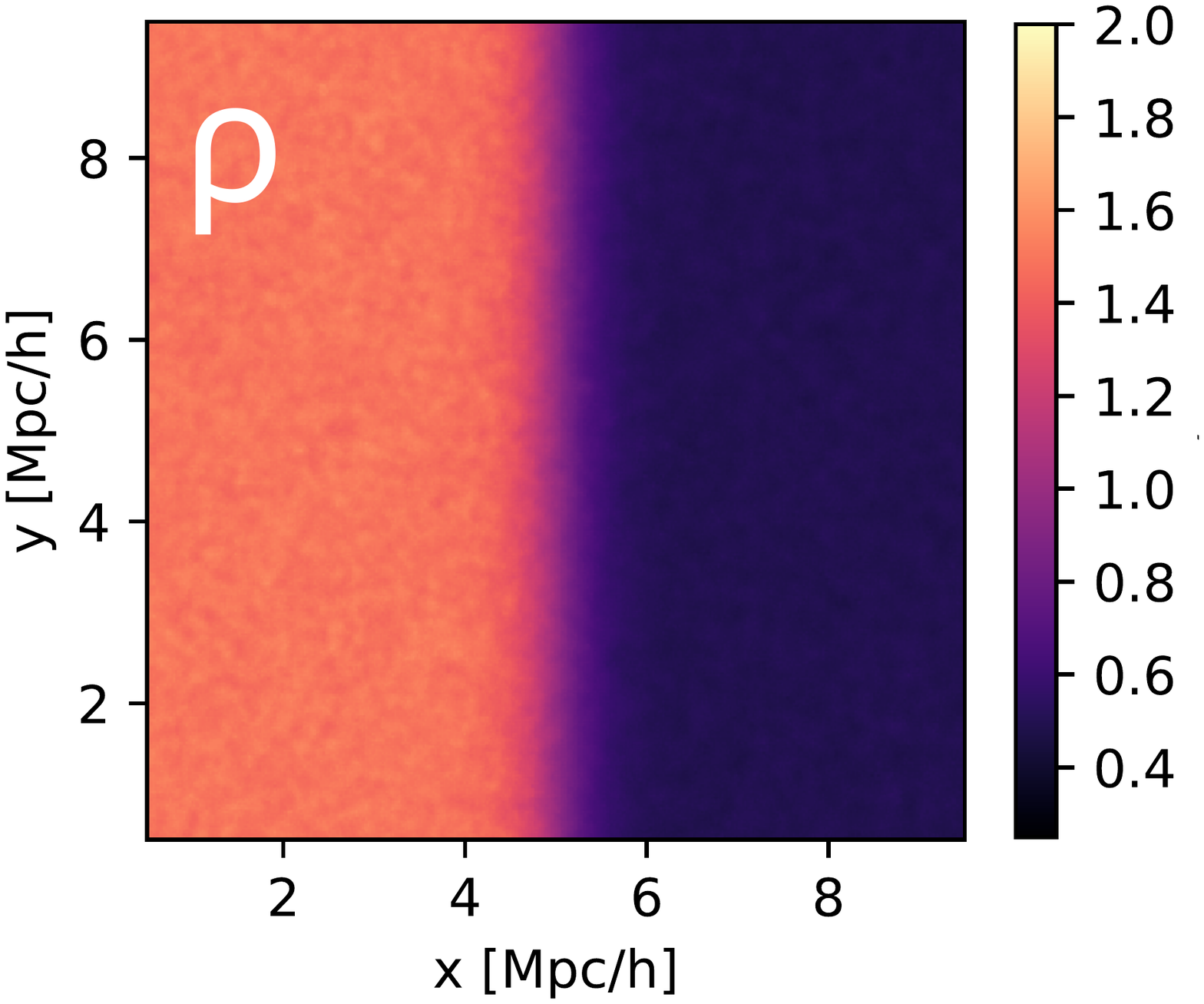}
\includegraphics[width=0.246\textwidth,trim={4.6cm 1.3cm 5.1cm 3.8cm},clip]{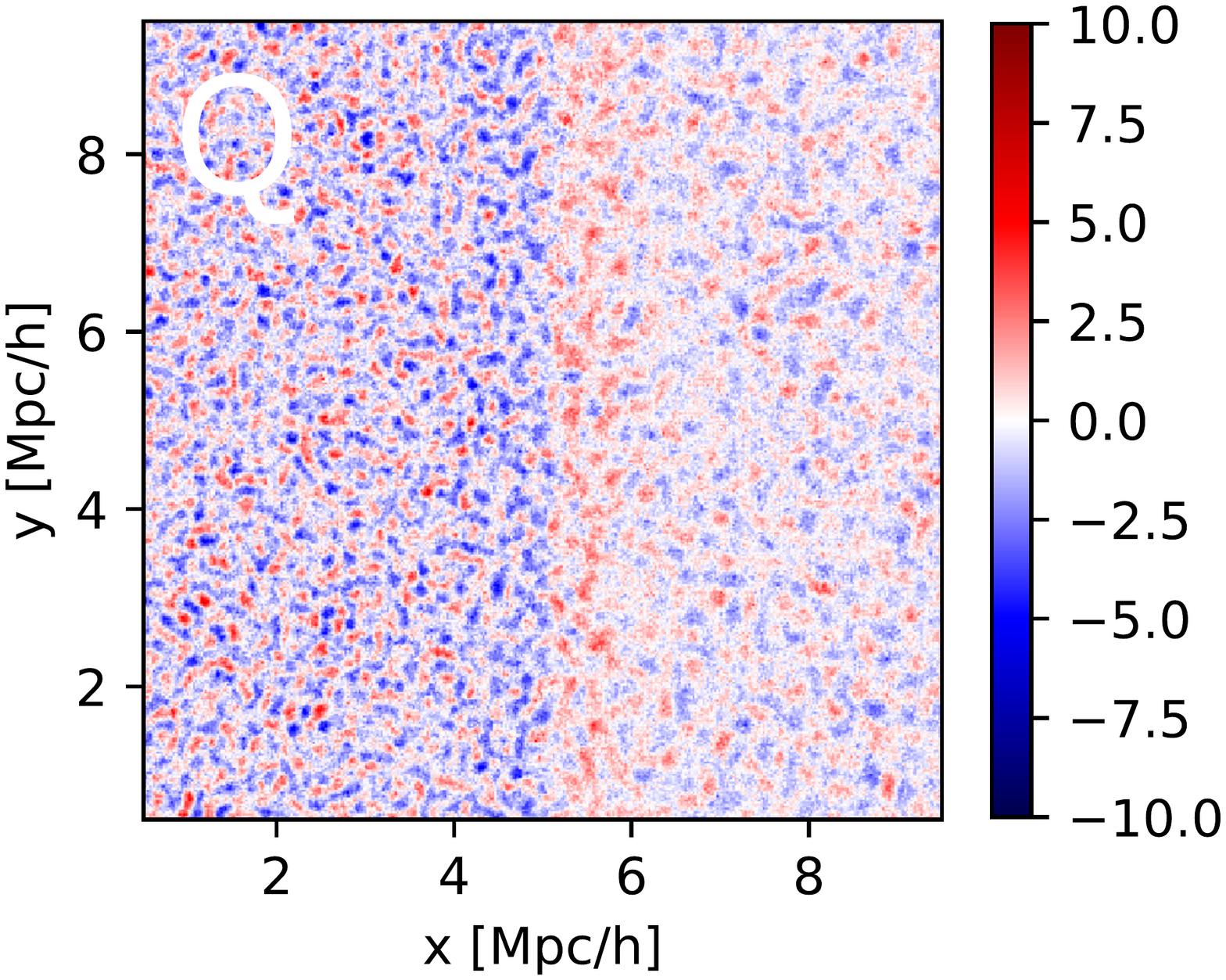}
\includegraphics[width=0.246\textwidth,trim={3.8cm 1.0cm 3.2cm 2.0cm},clip]{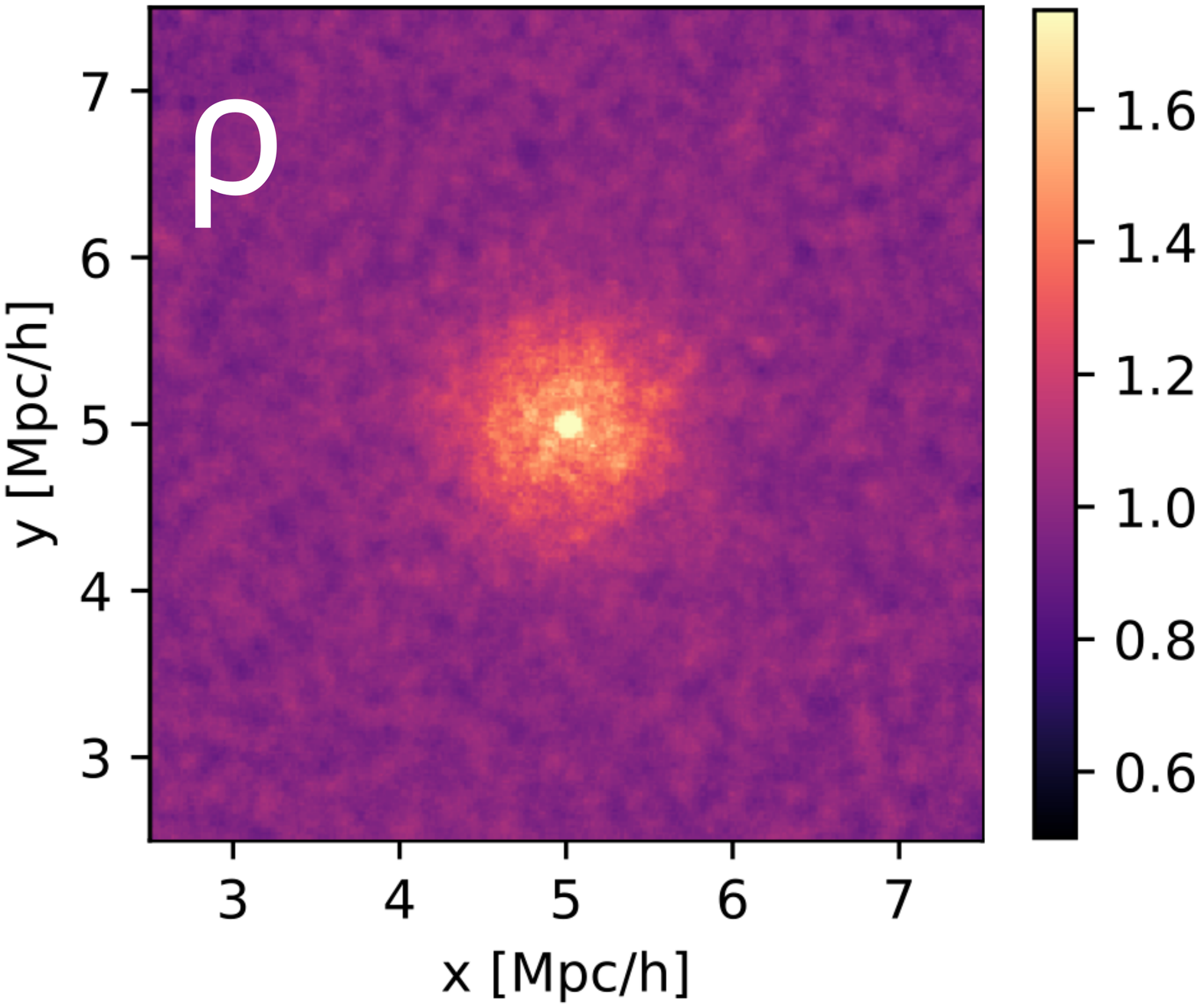}
\includegraphics[width=0.246\textwidth,trim={3.8cm 0.9cm 3.2cm 2.0cm},clip]{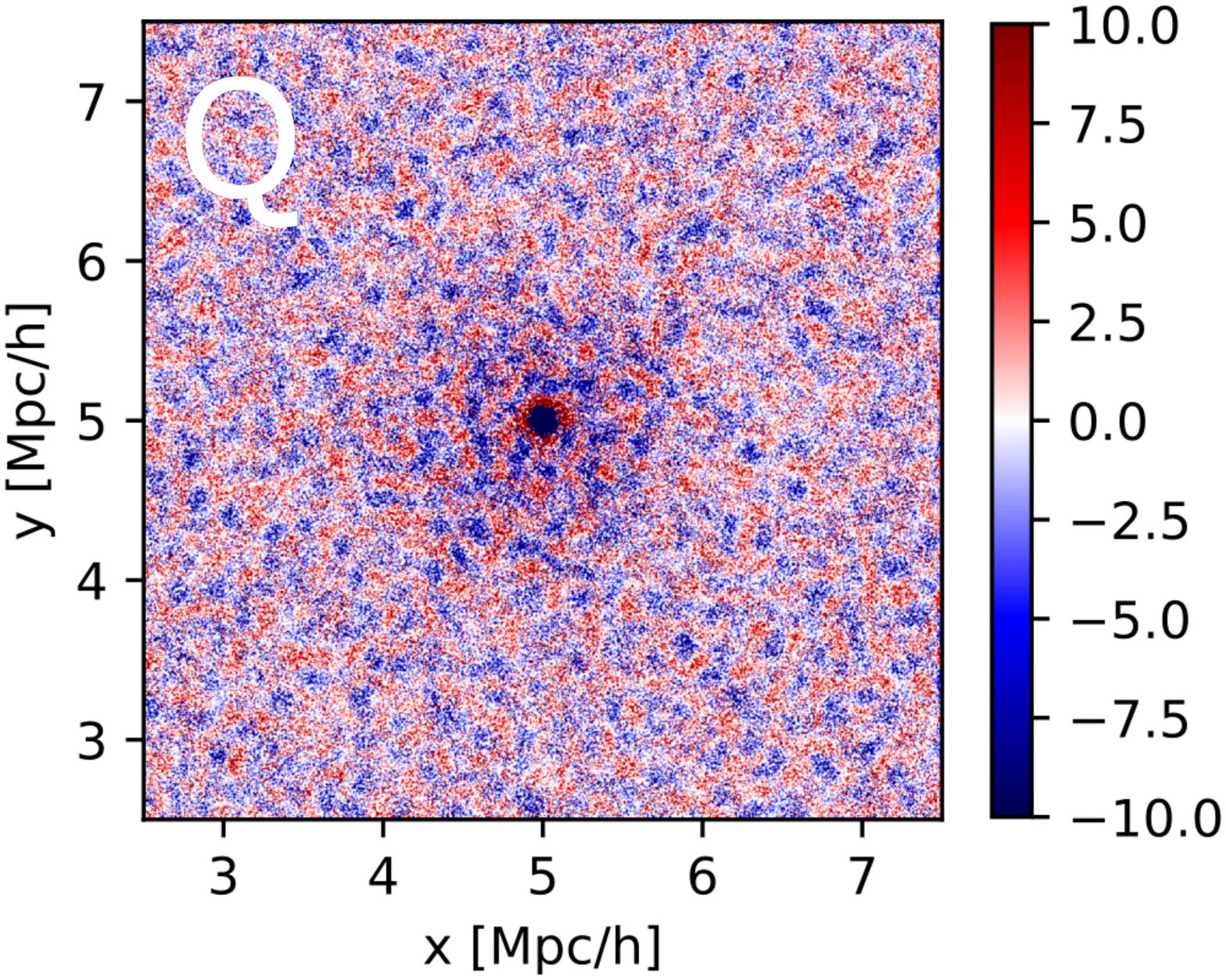}
\caption{Matter density and Quantum Potential maps obtained for the two analytical density distributions considered for code validation: a 1D hyperbolic tangent front along the $x$-axis ({\em left panels}) and a 3D Gaussian ({\em right panels}) distribution.}
\label{fig:TEST_MAPS}
\end{figure*}

To this end, we have tailored the matter density distribution in our tests, both in one and three dimensions, to match some specific analytical forms by rearranging the spatial distribution of particles while keeping their individual mass constant, which implies a local variation of the SPH smoothing length from particle to particle as it normally happens in standard astrophysical and cosmological application of the SPH method. The spatial degrees of freedom not relevant for the test distributions are uniformly randomized in order to average out their contribution to Q. In Fig.~\ref{fig:TEST_MAPS} we display four maps representing the matter density distribution and the QP spatial distribution for the first two analytical test considered -- a 1D hyperbolic tangent front and a 3D Gaussian matter distribution -- and described in detail in the following subsections.

The results obtained using a standard SPH algorithm through the original formulation of Eq.~\ref{eq:LSPHnocorrection} and our modified implementation of Eq.~\ref{eq:LSPH} are presented to emphasize the importance of derivative corrections in the algorithm.

To be thorough, and to assess the impact of variable smoothing length on the accuracy of the solution, we also show a comparison with results obtained from a homogeneous distribution of particles with spatially variable mass that reproduce the same overall density distribution. 

The test simulations feature $256^3$ particles in a $L=10$~Mpc non-periodic box. Initial conditions, built accordingly to each test, are read by the code and a snapshot with QP information is instantly produced.

Finally, our last test focuses on the dynamical evolution of a self-gravitating system, in which the balance between the opposite effects of the gravitational and quantum potentials lead to a stable solution and the formation of a solitonic core.

\subsection{1D Density Front}
\label{sec:front}

As a first analytical test, we consider the case of a 1D density front described by a hyperbolic tangent in the form of
\begin{equation}
\label{eq:front_density}
\rho(x,\sigma,c) = \rho_0 \left( c + 1 - {\tanh{\frac x \sigma}} \right)
\end{equation}
where $\sigma$ defines the sharpness of the front while $c$ is used to parametrize the the density contrast at the left of the front with respect to a background density on its right.

For such density profile, the QP has the analytical form
\begin{equation}
\label{eq:TANHQP}
Q (x,\sigma,c) = - \frac{\hbar^2 }{8m^2\sigma^2}
\frac {1-t^2}{(c+1-t)^2} \left[ 1 - 4t \left( c + 1 \right) + 3 t^2 \right] 
\end{equation}
expressed in terms of $t(x,\sigma)=\tanh{\frac x \sigma}$.
\begin{figure*}
\centering
\includegraphics[width=\textwidth,trim={1.6cm 0.1cm 2.6cm 1.6cm},clip]{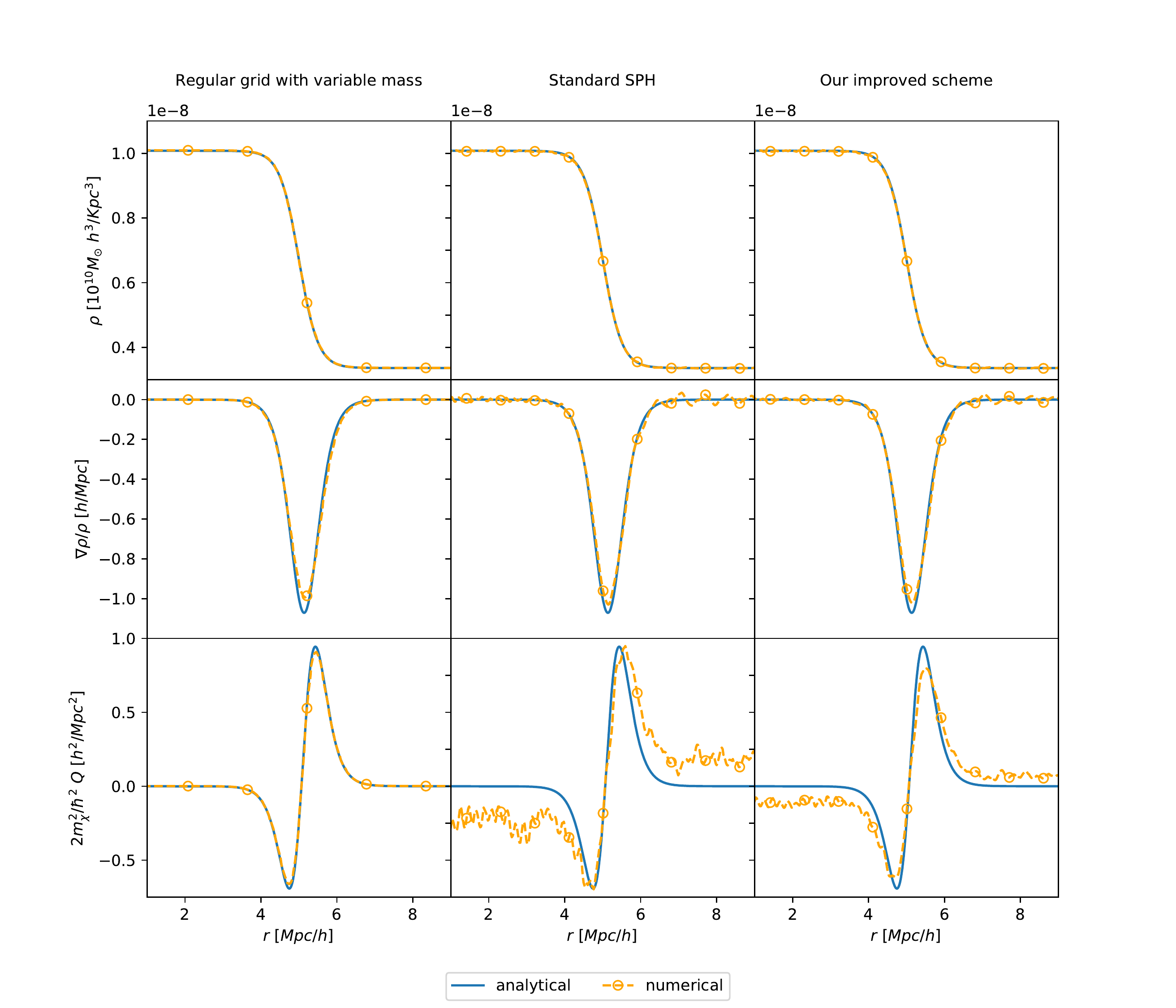}
\caption{Density profile ({\em top row}), Density gradient ({\em mid row}), and Quantum Potential ({\em bottom row}) obtained for the same hyperbolic tangent front density distribution along the $x$ axis. On the {\em left} column we show results obtained when the density distribution is built by changing the mass of particles set on a regular grid, while in the {\em center} and {\em right} columns we show the case of keeping the mass constant and rearranging the spatial distribution of particles, using either the original \G SPH scheme or our improved implementation, respectively.}
\label{fig:TANH_OBS}
\end{figure*}

In Fig.~\ref{fig:TANH_OBS} we show the profile for the density, its gradient, and the QP as computed for such density distribution in different setups: a regular grid configuration of particles with variable mass {(\em left panels)} and a spatial rearrangement of constant mass particles, the latter analysed with  and without the derivative correction described in section~\ref{sec:code} above ({\em center} and {\em right} panels, respectively). In particular, the distribution used has the form of Eq.~\ref{eq:front_density} centered at $5$~Mpc, with $\sigma=500$~kpc and a background parameter $c=1$.

First of all, we notice how the idealised setup with a regular grid of particles with variable mass provides the most accurate solution for all quantities. This is not surprising, and reflects an intrinsic limitation of the SPH algorithm in the computation of spatial derivatives for situations where the density distribution features steep gradients and consequently neighbouring particles have significantly different smoothing lengths. By keeping particles on a fixed grid and changing their mass -- still basing the computation of the SPH smoothing length on a desired number of neighbour particles -- we obtained identical smoothing lengths for all particles.

Secondly, it is easy to see that our correction of the derivative scheme has a positive impact on the QP computation which gets closer to the idealized variable mass system.

\begin{figure}
\includegraphics[width=\columnwidth,trim={1.5cm 0.3cm 1.6cm 1.4cm},clip]{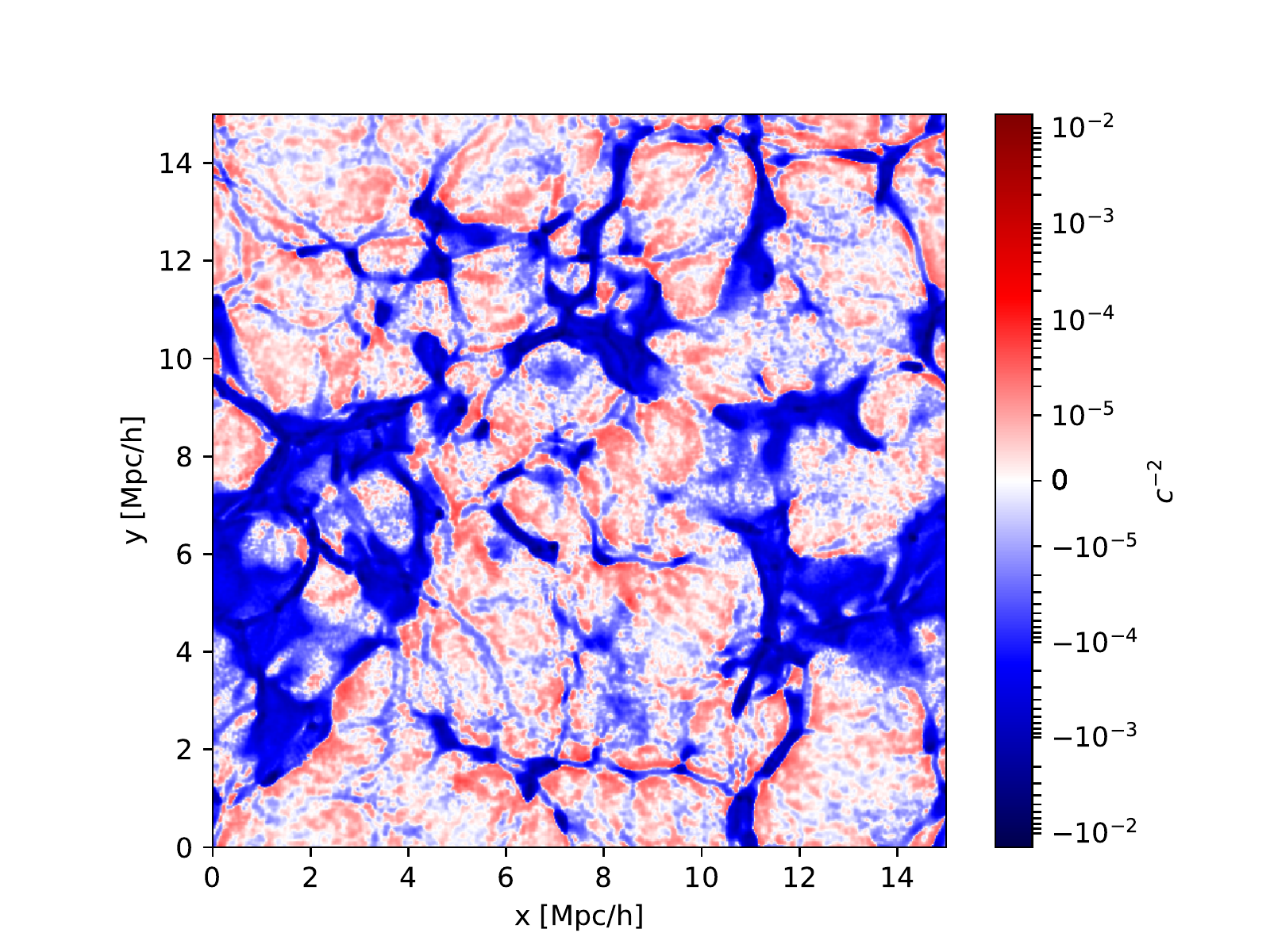}
\caption{Quantum Potential map at $z=5$ of a FDM cosmological simulation  with $m_\chi=10^{-22}$ eV$/c^2$ in a $15$ Mpc$/h$ side box. The potential is contrasted with its mean value to emphasize spatial distribution, therefore expressed in dimensionless units.}
\label{fig:TANH_filaments}
\end{figure}

As one can see in the plots, the QP resulting from a 1D hyperbolic tangent front features a negative peak on the most dense side and a positive one towards the less dense region. This corresponds to positive and negative accelerations on the two sides, respectively, implying that the QP tends to push matter towards the region of steep density variation. Such behaviour has interesting consequences, in particular for cosmological structure formation. More specifically, this modulation of the QP may show up in cosmic walls and filaments, where the local matter distribution can be represented by a 1D (cartesian or radial, respectively) density front. In order to provide a direct evidence of such effect, we show in Fig.~\ref{fig:TANH_filaments} a map of the QP contrast (i.e. the relative difference of the QP to the average QP) at redshift $z=5$ from a cosmological simulation of Axion Dark Matter performed with \AG for an Axion mass of $m_\chi= 10^{-22}$~eV$/c^2$, where negative and positive regions for the QP are marked in blue and red, respectively. As one can see from the figure, the QP follows the underlying cosmic web of collapsed structures, and shows negative wells corresponding to the most dense regions while approaches zero in voids, as expected. It is also clearly visible how positive regions surround structures -- filaments and walls in particular -- that separate voids, confirming the analytical result obtained above.

\subsection{3D Gaussian distribution}
\label{sec:gauss}

To idealize a spherical collapsed system we used as pivotal test a 3D Gaussian overdensity, that we parametrized as
\begin{equation}
\label{eq:Gaussian_density}
\rho(r,\sigma,c) = \rho_0 \left( c + e^{-r^2/2\sigma^2} \right)
\end{equation}
where $\sigma$ is the standard deviation of the distribution and $c$ is linked to the relative density of the Gaussian perturbation with respect to the background average.

For $c\rightarrow 0$, representing the Gaussian density distribution in vacuum space, $Q(x,\sigma,0)$ collapse into a parabolic function that diverges at infinite distance. This implies a unphysical limit in which distant particles have infinite acceleration (proportional to $\vec \nabla Q$).

In fact, using Eq.~\ref{eq:QP} the QP functional form is
\begin{equation}
\label{eq:GAUSSQP}
Q (r,\sigma,c) = \frac{\hbar^2}{4m^2\sigma^2} \ \chi \left[ - 3 + \frac{r^2}{2\sigma^2} \left(2-\chi \right) \right]
\end{equation}
where we introduced the dimensionless variable $\chi (r,\sigma) = ( 1 + c \  \exp(r^2/2\sigma^2) )^{-1}$.

As soon as $c$ becomes different from zero, the divergence of the QP is cured and a positive peak appears outside the central negative well, similarly to the previous case, before the function decays to zero at larger distances. The acceleration
\begin{equation}
\vec \nabla Q(r,\sigma,c) = 
\frac{\hbar^2 }{4m^2\sigma^3} \ \chi \left[ 5 - 4 \chi - \frac {r^2} {\sigma^2} \left( 1 - \chi \right)^2 \right] \frac {\vec r} {\sigma}
\end{equation}
is therefore directed outwards in the central overdensity region and inwards in a small shell in the outskirts of the overdensity.

This non-linear behaviour in a simple Gaussian distribution is emblematic of how the QP is hardly representable with other effective functionals such, for example, a polythropic function $Q \propto \rho^{\gamma}$ typical of pressure-like components that would feature a monotonic behaviour (i.e. with an acceleration with fixed sign) whatever its specific form.

\begin{figure*}
\centering
\includegraphics[width=\textwidth,trim={1.6cm 0.1cm 2.6cm 1.6cm},clip]{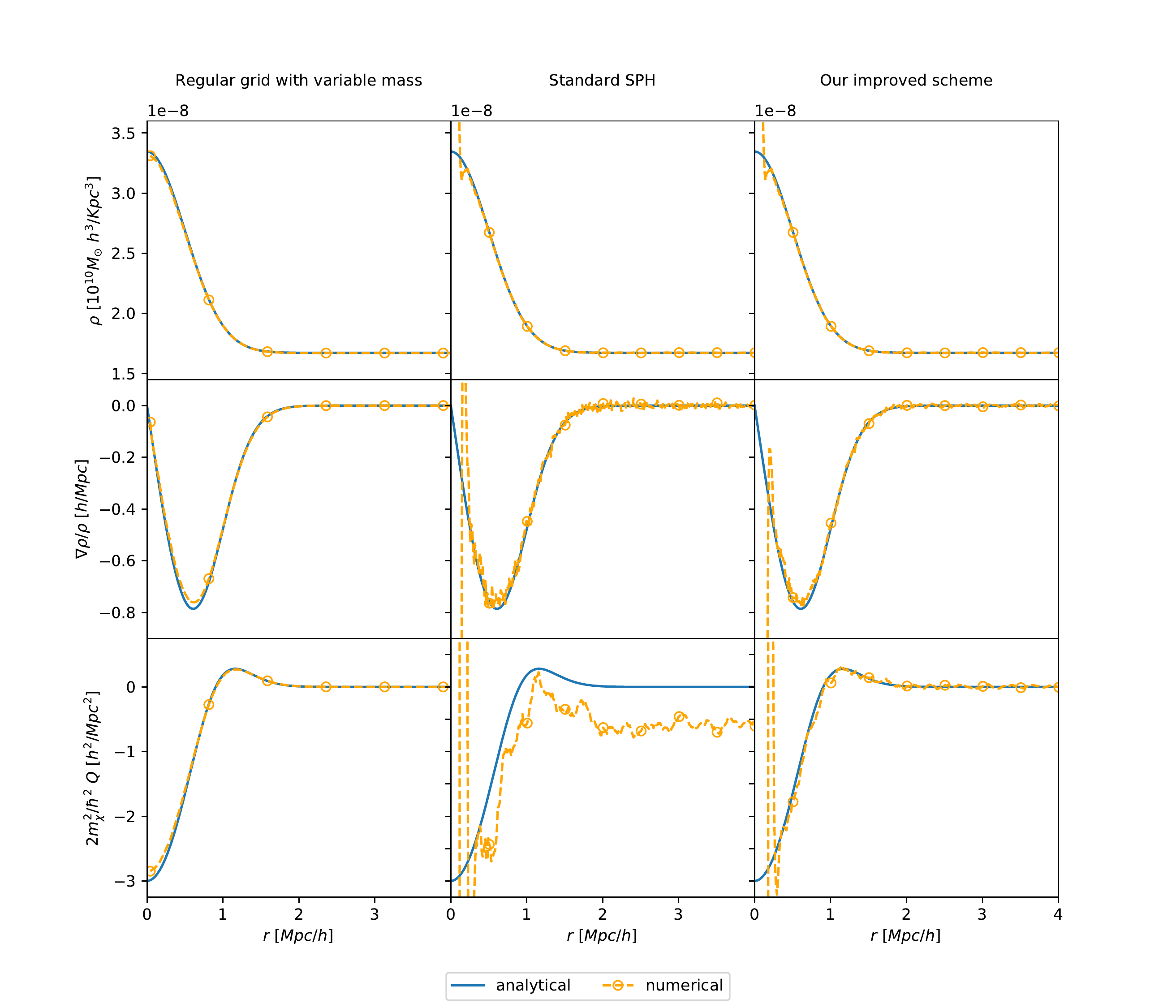}
\caption{Density profile ({\em top row}), Density gradient ({\em mid row}), and Quantum Potential ({\em bottom row}) 
obtained for the same 3D Gaussian density distribution. On the {\em left} column we show results obtained when the density distribution is built by changing the mass of particles set on a regular grid, while in the {\em center} and {\em right} columns we show the case of keeping the mass constant and rearranging the spatial distribution of particles, using either the original \G SPH scheme or our improved implementation, respectively.}
\label{fig:GAUSS_3D_OBS}
\end{figure*}

In Fig.~\ref{fig:GAUSS_3D_OBS} we display the density profile, its gradient, and the resulting QP for the 3D Gaussian distribution of Eq.~\ref{eq:Gaussian_density} around the center of a $10$~Mpc non-periodic box, with $\sigma=500$~ kpc and a background parameter $c=1$.

As the figure shows, the role of our derivative correction is relevant, since the standard SPH approach results in an overestimation of the depth of the central well lacking of positive peaks surrounding it. From the numerical point of view, our analysis suggests that the main source of error comes from a substantial underestimation of the Laplacian, since it is the only term bearing a positive contribution.

Recently, another implementation for FDM from an independent group \citep{Zhang16} considered a similar test, finding the opposite behaviour, namely that the interaction induced by the QP between two particles is attractive below a certain distance and repulsive elsewhere. Clearly such solution would bound particles that are close enough to each other and was claimed by \citeauthor{Zhang16} to reflect the phenomenon of Bose-Einstein condensation. However, it is our opinion that the attractive behaviour found by the latter could be actually due to a mis-calculation in the discretization of the SPH algorithm. We provide a more detailed demonstration of this argument and a comment on the results of \citeauthor{Zhang16} in Appendix~\ref{sec:zhang} below. 

\subsection{Solitonic core}
\label{sec:soliton}

The last test we present features the dynamical evolution of an analytical distribution, in order to test the correctness of our implementation of FDM dynamics over time. The starting point is again a 3D Gaussian distribution 
\begin{equation}
\rho(r,\sigma) = \rho_0 \ e^{-r^2/2\sigma^2}
\end{equation}
which is left free to evolve under the influence of both gravitational and quantum potential, in a non-cosmological setup.

The stable solution of the density distribution for this system -- representing a {\it solitonic} solution -- has no analytical form but can be expressed in an approximated form as
\begin{equation}
\label{eq:soliton}
\rho(r,\sigma) \xrightarrow{t \rightarrow \infty} \rho(r,r_c) = \rho_c \left[ 1 + \alpha r^2/r_c^2 \right]^{-8}
\end{equation}
where the parameter $\alpha = \sqrt[8]{2} - 1$ is defined such that the the radius $r_c$ is the radius at which the density is halved with respect to the central peak $\rho_c$ satisfying $\rho(r=r_c) = \rho_c / 2$ \citep[see e.g.][]{Guzman06,Schive14,Marsh15CCP}.
\begin{figure}
\includegraphics[width=\columnwidth,trim={0.8cm 0.3cm 1.3cm 1.0cm},clip]{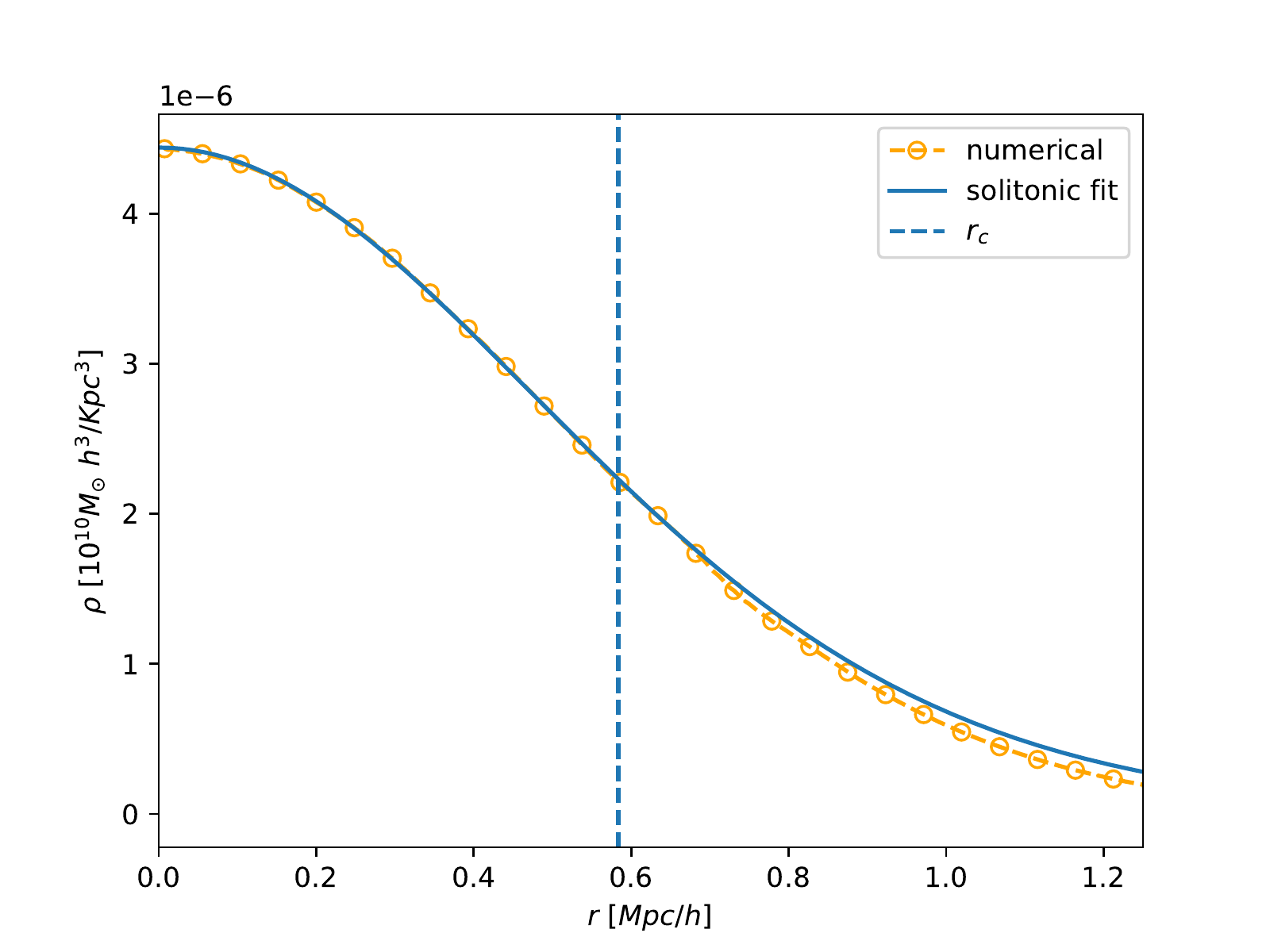}
\caption{Stable density distribution obtained by letting a 3D Gaussian distribution relax over time under the combined effect of gravitational and quantum potentials. The numerical result matches the solitonic core functional form of Eq.~\ref{eq:soliton}, which parameter $r_c$ is represented by the vertical dashed line.}
\label{fig:SOLITON}
\end{figure}

We choose a non-periodic box of $10$~Mpc side length, with a Gaussian distribution with $\sigma=500$~ kpc and a mass $m_\chi = 10^{-26}$ eV$/c^2$ for the FDM bosonic field.

In Fig.~\ref{fig:SOLITON} is shown the relaxed radial density distribution obtained for such setup, which is consistent with the approximated solitonic functional form of Eq.~\ref{eq:soliton} with $r_c$ represented by the vertical dashed line. In this idealized test, we added an artificial velocity dependent friction in order to let the system settle in its ground state more gracefully and achieve stability.

This last test is in line with the theoretical predictions and numerical results in literature, assessing that the QP can indeed support the formation of stable and cored structures \citep[][]{Schive14,Marsh15CCP}. In a cosmological setup, since the scale of equivalence between the two potentials $\lambda_Q$ evolves in time, such solitonic cores are expected to be found only at the center of small and dense dark matter haloes which had enough time to relax dynamically.

Therefore, by comparing our results with both analytical predictions and numerical results in the literature \citep[see e.g. ][]{Woo09,Mocz18} for static and evolving systems, we feel confident that the algorithm implemented in \AG can be considered accurate and robust, and we now move on to test the effects of FDM with our modified code on more realistic cosmological setups.

\section{Cosmological simulations}
\label{sec:sims}

In this section we discuss the results of a series of cosmological test simulations, to understand the effects of the QP on the overall dynamics of FDM and its role in the evolution of the large-scale structures of the universe.

Even though \AG allows for any possible mixture of CDM and FDM particles sharing the overall matter budget of the universe, as well as for different possible self-interaction mechanisms as described in section~\ref{sec:code}, we restrict our tests to the effects of the QP alone for a single FDM component accounting for the total dark matter density and leave the exploration of more complex models for future work.

The dynamical effect induced by the QP during cosmic evolution is investigated and compared to the result of the widely-adopted approximation consisting in imposing the predicted suppression of small-scale perturbations -- as computed by linear codes such as e.g. {\small axion-CAMB} \citep[][]{axionCAMB} -- in the initial conditions only. The details regarding each simulation are listed in Table~\ref{tab:SIMS}.

In order to highlight the effects of the QP on structure formation and -- more importantly -- the ability of AX-Gadget to correctly follow FDM dynamics, we choose a mass of the FDM boson field $m_\chi \leq 10^{-22}$ eV$/c^2$. A single FDM component with such low mass is disfavoured by linear studies \cite{axionCAMB} and by numerical simulations based on a suppressed initial density power \citep[][]{Irsic17}. However, in the present work we are mostly interested in testing our code in the case of a strong QP effect in order to emphasise observable consequences of FDM on cosmological evolution and numerically stress the code.

In these simulation we are not able to see the formation of interference patterns or solitonic cores -- as in section 3.3 -- since we are not able to probe such high-resolution effects. With much higher resolution future simulations we plan to investigate the ability of our SPH scheme to capture such small scales characteristic FDM footprint and large scale structures at the same time, as has been done with other grid-based codes in the literature \citep{Schive14,Mocz18}.

\begin{table*}
\centering
\caption{Summary of the simulations presented in this work and their properties.}
\label{tab:SIMS}
\begin{tabular}{cccccccccc}
\hline
\multirow{2}{*}{Model} & \multirow{2}{*}{Initial Contitions} & \multirow{2}{*}{QP} & $m_{\chi}$          & \multirow{2}{*}{$N$ particles} & Boxsize   & Mass resolution          & \multirow{2}{*}{$z_\text{start}$} & \multirow{2}{*}{$z_\text{end}$} & Time                    \\
                       &                                     &                     & $[10^{-22} eV/c^2]$ &                                & $[Mpc^3]$ & $[10^{6} M_\odot]$ &                                   &                                 & $[h~\text{on}~1024~\text{CPU}]$ \\ \hline
CIC$1000$              & standard                            & $\times$            & -                   & $256^3$                        & $10$      & $5.110$            & $999$                             & $0.5$                           & -                       \\
CIC$+$QP$1000$         & standard                            & $\checkmark$        & $1/\sqrt{2}$        & $256^3$                        & $10$      & $5.110$            & $999$                             & $0.5$                           & -                       \\ \hline
CIC                    & standard                            & $\times$            & -                   & $512^3$                        & $15$      & $2.156$            & $99$                              & $3$                             & 6.98                    \\
CIC$+$QP               & standard                            & $\checkmark$        & $1$        & $512^3$                        & $15$      & $2.156$            & $99$                              & $3$                             & 38.09                   \\
FIC                    & suppressed                          & $\times$            & -                   & $512^3$                        & $15$      & $2.156$            & $99$                              & $3$                             & 7.29                    \\
FIC$+$QP               & suppressed                          & $\checkmark$        & $1$        & $512^3$                        & $15$      & $2.156$            & $99$                              & $3$                             & 38.27                   \\ \hline
\end{tabular}
\end{table*}

\subsection{Quantum Potential effects on dynamics}

To isolate the impact of the QP on the dynamics and on the evolution of large-scale structures, we first performed two simulations -- termed CIC$1000$ and CIC$+$QP$1000$ in Table~\ref{tab:SIMS} --  evolved with CDM and FDM dynamics (i.e. with standard \G and \AG, respectively) but starting from the same initial conditions. Therefore, any difference in the dynamical evolution between the two runs are the result of the QP acceleration contribution exclusively. The initial power spectrum used for the initial particle configuration is that of $\Lambda$CDM, specifically the Eisenstein and Hu spectrum \citep{Eisenstein97}, in order to avoid additional effects arising from the suppression that a light non-thermal boson field would imply \citep{Hu00}. 

Since the initial conditions that we adopt feature a higher power of density perturbations at small scales than FDM would allow, we set the starting time of the simulation at a very high redshift $z=999$ in order to allow sufficient time for the the system to adjust. Such approach has been already employed to quantify the QP effects on structure formation for full-wave solver codes \citep[see e.g.][]{Woo09} and our tests therefore allow for a direct comparison with these previous works.

\begin{figure*}
\includegraphics[width=\textwidth, trim={0cm 2cm 0cm 2.5cm},clip ]{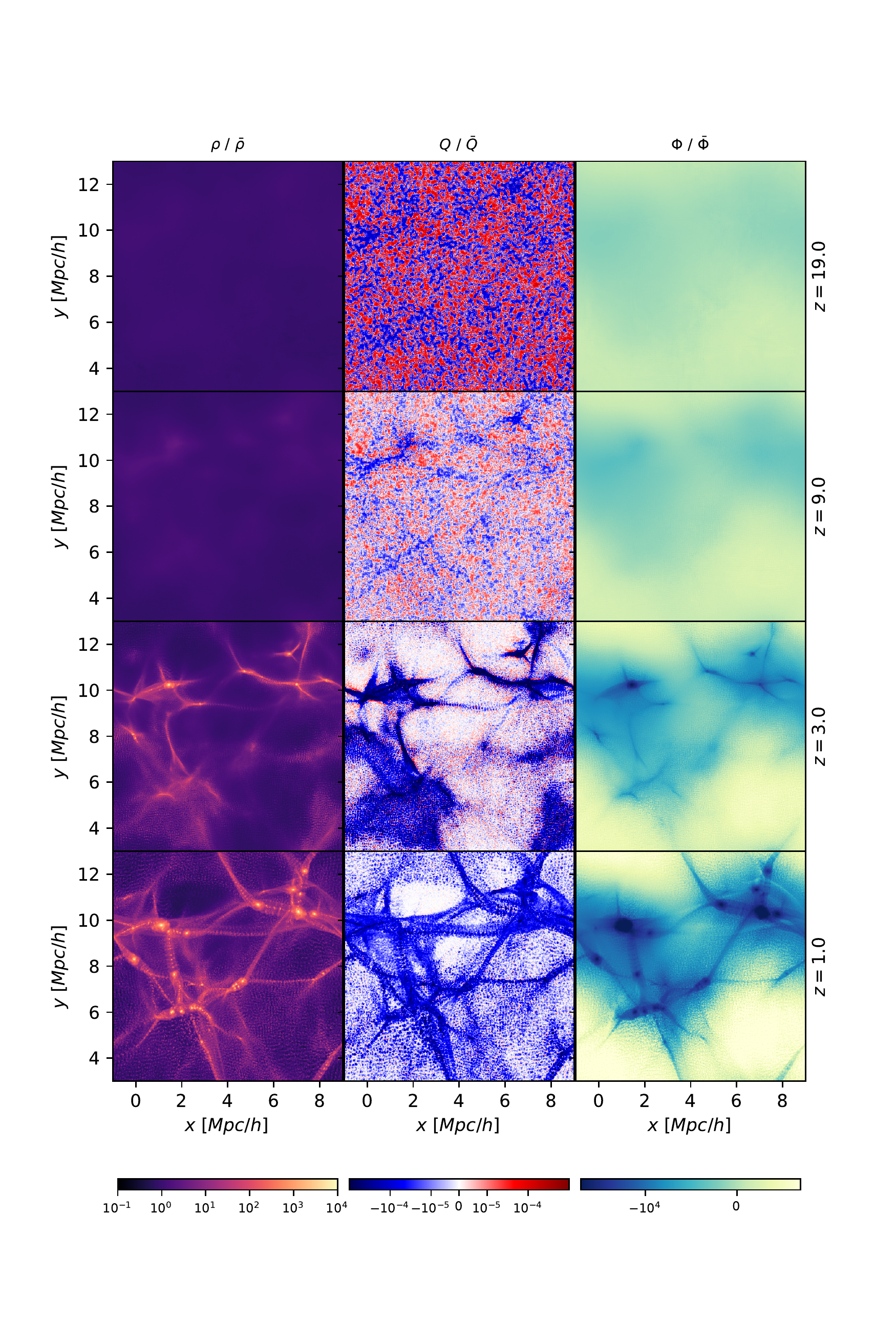}
     \caption{Maps of a $500$ kpc$/h$ slice of the density field (\textit{left column}), the Quantum Potential (\textit{center column}) and the gravitational potential (\textit{right column}) of the FDM simulation, at different redshifts. Observables are contrasted with mean values to emphasize spatial distribution, therefore expressed in dimensionless units.}
\label{fig:MAPS}
\end{figure*}

\begin{figure}
\includegraphics[width=\columnwidth,trim={0cm 1.0cm 1.6cm 2.3cm},clip]{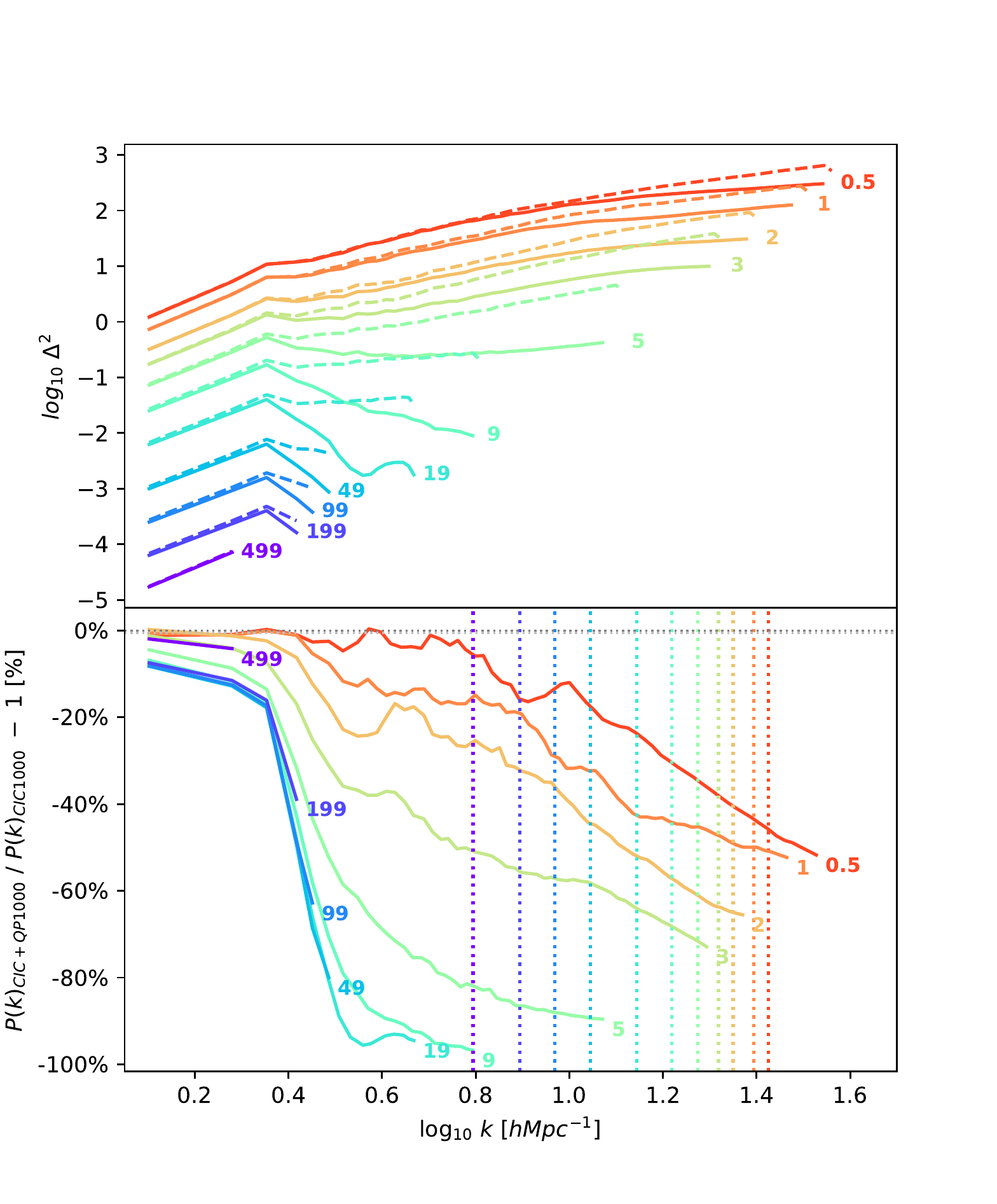}
\caption{\textit{Top:} Dimensionless matter power spectra of simulations with QP (solid) and without QP (dashed), both evolved from $z=999$ CDM initial condition, labelled redshift-wise. \textit{Bottom:} Relative difference of matter power spectra between the simulations. The vertical dashed lines represents the $k_{1/2}$ coming from linear theory defined as the scale at which the predicted power spectrum suppression is $50\%$.}
\label{fig:PS1000}
\end{figure}

In Fig.~\ref{fig:MAPS} we display maps of the density field (\textit{left column}), the QP (\textit{central column}) and the gravitational potential (\textit{right column}) of the CIC$+$QP$1000$ simulation at different redshifts. 
At high redshift, the density contrast is still very small, and the QP is strongly affecting scales $k\lesssim 1~h/$Mpc throughout the simulation box -- thereby counteracting gravitational instability at these scales --  while the gravitational potential wells start to induce matter collapse on larger scales.
As the system evolves, dark matter starts to accrete on seed overdensities under the effect of gravity and eventually collapses into structures while drifting away from low density regions, thus inducing the QP to intensify in the infalling regions -- actively counteracting matter accretion -- and to weaken its action elsewhere.
At lower redshifts the scale at which the QP is still able to contrast the gravitational potential reduces -- because of the redshift dependence of the associated Jeans scale of Eq.~\ref{eq:kq} -- and its distribution follows the dark matter structures shaped by the gravitational potential.

In Fig.~\ref{fig:PS1000} we show the matter power spectra of both simulations (in the {\em upper panel}) and their ratio (in the {\em lower panel}).
It appears clear from the evolution of the power spectra that the QP dynamically suppresses the power at small scales, as expected. Both the intensity and the scale of this suppression are redshift dependent and the evolution of the system can be split in three main phases.
In the first phase, at high redshifts (up to $z\sim19$), the QP fluctuations dominate over the gravitational potential at small scales,resulting in a strong suppression of the initial power spectrum.

As the the system evolves, it enters a second phase (from $z\sim19$ to $z\sim3$ in our plots) where the action range of the QP gets smaller and smaller (as described by its characteristic length $k_{Q}$ in Eq.~\ref{eq:kq}) and it is no longer able to counteract the gravitational potential on larger scales. Structures begin to form at scales larger than $k_{Q}$ as the gravitational instability induces matter collapse, thereby increasing density gradients in the collapsing regions and consequently intensifying the repulsive action of the QP in the central parts of the forming halos.

Such first dramatic rearrangement of the initial conditions followed by a smoother evolution, in line with what is found in \citet{Woo09}, is due to the fact that the CDM initial conditions are not an equilibrium solution in the presence of QP, so that the system suddenly rearranges to recover an equilibrium setup. Unfortunately, the significant difference in resolution between our runs and those of \citeauthor{Woo09} prevents a detailed quantitative comparison of this sharp transition between the two studies.

Finally, at even lower redshifts (from $z\sim3$ onward) gravity has shaped the large scale structures and both potentials - effectively acting one against the other -  follow the matter distribution and relax to an equilibrium state. 

The suppression of the power spectrum -- displayed in Fig.~\ref{fig:PS1000} -- shows no dramatic change in slope while it shifts towards lower and lower scales, suggesting that the evolution of perturbations is mostly due to the dynamical balance at all scales of the two potentials as the universe expands and the quantum Jeans length shrinks.

As a test of the dynamical evolution of the system in the N-body simulation, we can compare our results with with the linear prediction for $k_{1/2}$, which is defined as the scale satisfying $P_{\rm FDM}(k_{1/2})/P_{\rm CDM}(k_{1/2}) -1 = -50\% $ \citep[see e.g.][]{Hu00}. In the bottom panel of Fig.~\ref{fig:PS1000} the vertical dashed lines represent $k_{1/2}$ for each redshift.

In the first phases of the simulation -- when the system quickly shocks from the non-equilibrium configuration of the initial conditions -- the linear prediction is far from being realised, while in the last phase, when sufficient time has been allowed for the system to settle to the new equilibrium configuration, we progressively approach the linear result as the universe evolves to the present epoch. Such asymptotic recovery of the linear predictions at low redshifts represents a successful test for our QP implementation: given enough time, the QP is able to prevent structure formation at small scales -- even when CDM initial conditions are used -- effectively suppressing the matter power spectrum as we would expect from theory. Nonetheless, we stress here that we do not expect to recover exactly the predicted value of $k_{1/2}$ at low redshift as the latter was computed within a linear approximation while our implementation is able to follow the evolution of structures under the joint effects of gravity and of the QP down to the fully non-linear regime. The differences between the linear results and our non-linear treatment is discussed in detail below.

\subsection{Quantum Potential and initial conditions}

As we showed in the previous subsection, comparing the evolution of CDM initial conditions with and without accounting for QP in the dynamics provides a clear example of the QP effect on structure formation and evolution, namely a repulsive contribution to acceleration within collapsed structures that counteracts the attractive pull of gravity. However, such setup is not a realistic representation of structure formation within the FDM framework, for which a suppression of the density perturbations power at small scales would already be in place at arbitrarily high redshifts, and therefore should be already accounted for in the initial conditions setup \citep[see again][]{Hu00}.

A lower small-scale power translates automatically into a late-time shortage of low-mass structures, so different works \citep[see e.g.][]{Schive16,Armengaud17,Irsic17} have been suggesting that it might be appropriate -- under some circumstances -- to completely neglect the effects of the QP in the dynamics of the simulations and simply account for the FDM phenomenology through a cutoff in the initial conditions power spectrum, similarly to what happens for the case of Warm Dark Matter scenarios \citep[see e.g.][]{Bode00}. However, a proper validation of such approach has not been yet performed in sufficient detail, and a quantitative assessment of the impact of the QP in the dynamics of structure formation on top of a cutoff in the primordial power spectrum has to be made in order to allow fully accurate predictions of the non-linear FDM footprints.

To this end, in order to investigate the relative impact on structure formation of the two approaches, we performed a second set of four simulations -- termed CIC, CIC$+$QP, FIC, FIC$+$QP -- representing four combinations obtained from both suppressed (FIC) and non-suppressed (CIC) initial conditions evolved either with ($+$QP) or without the QP in the dynamics, as listed in Tab.~\ref{tab:SIMS}. 

Obviously, the CIC run corresponds to a standard $\Lambda$CDM simulation while FIC$+$QP represents the closest setup to the real FDM model, including the effects of the QP both in the initial conditions and in the subsequent dynamical evolution of structures.

As already pointed out in Section \ref{sec:maths}, the matter power spectrum to be taken into account in building the initial conditions for FDM simulations features a mass- and redshift-dependent cutoff, at a scale given by Eq.~\ref{eq:kq}. In order to set up the FIC initial conditions, we resorted on the publicly available and widely used code {\small axionCAMB} \citep{axionCAMB}, a modified version of the public code {\small CAMB} \citep[][]{CAMB}, to compute the suppressed power spectrum at the starting redshift of our simulations, $z=99$. 

\begin{figure*}
\begin{tabular}{cc}
$z=99$ & $z=49$ \\
\includegraphics[width=\columnwidth,trim={0cm 0cm 1.5cm 1cm},clip]{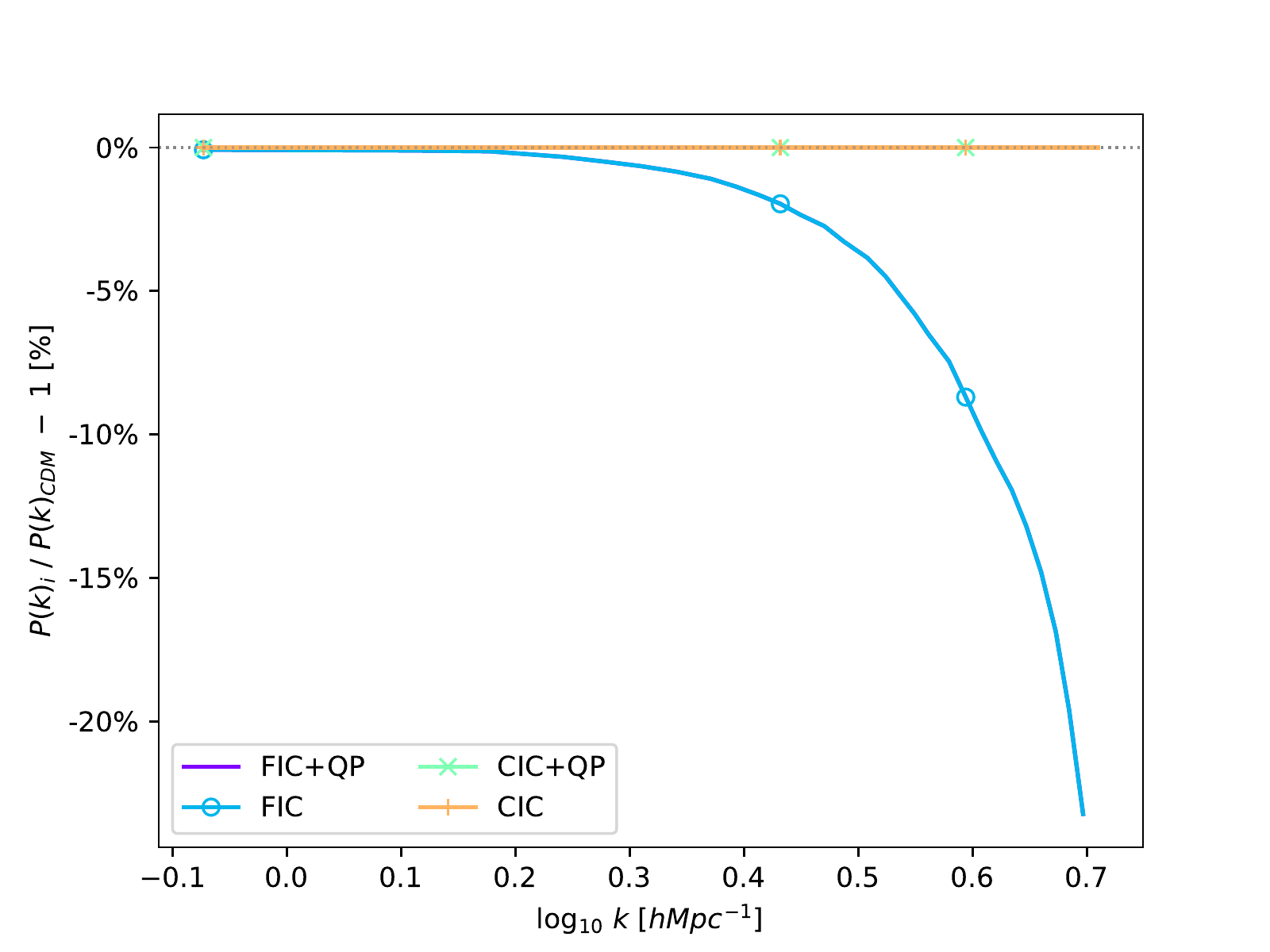} &
\includegraphics[width=\columnwidth,trim={0cm 0cm 1.5cm 1cm},clip]{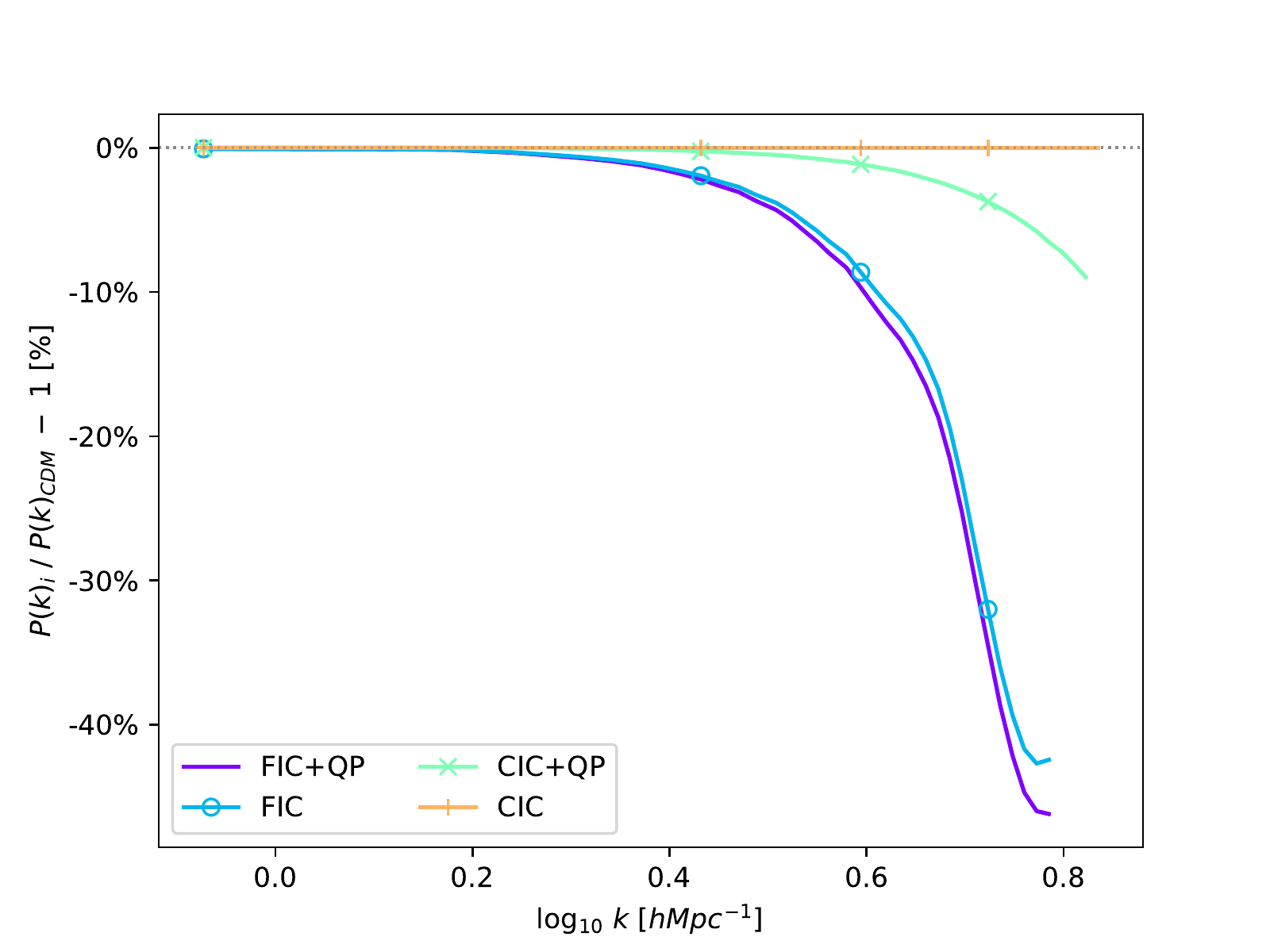} \\
\\
$z=9$ & $z=3$ \\
\includegraphics[width=\columnwidth,trim={0cm 0cm 1.5cm 1cm},clip]{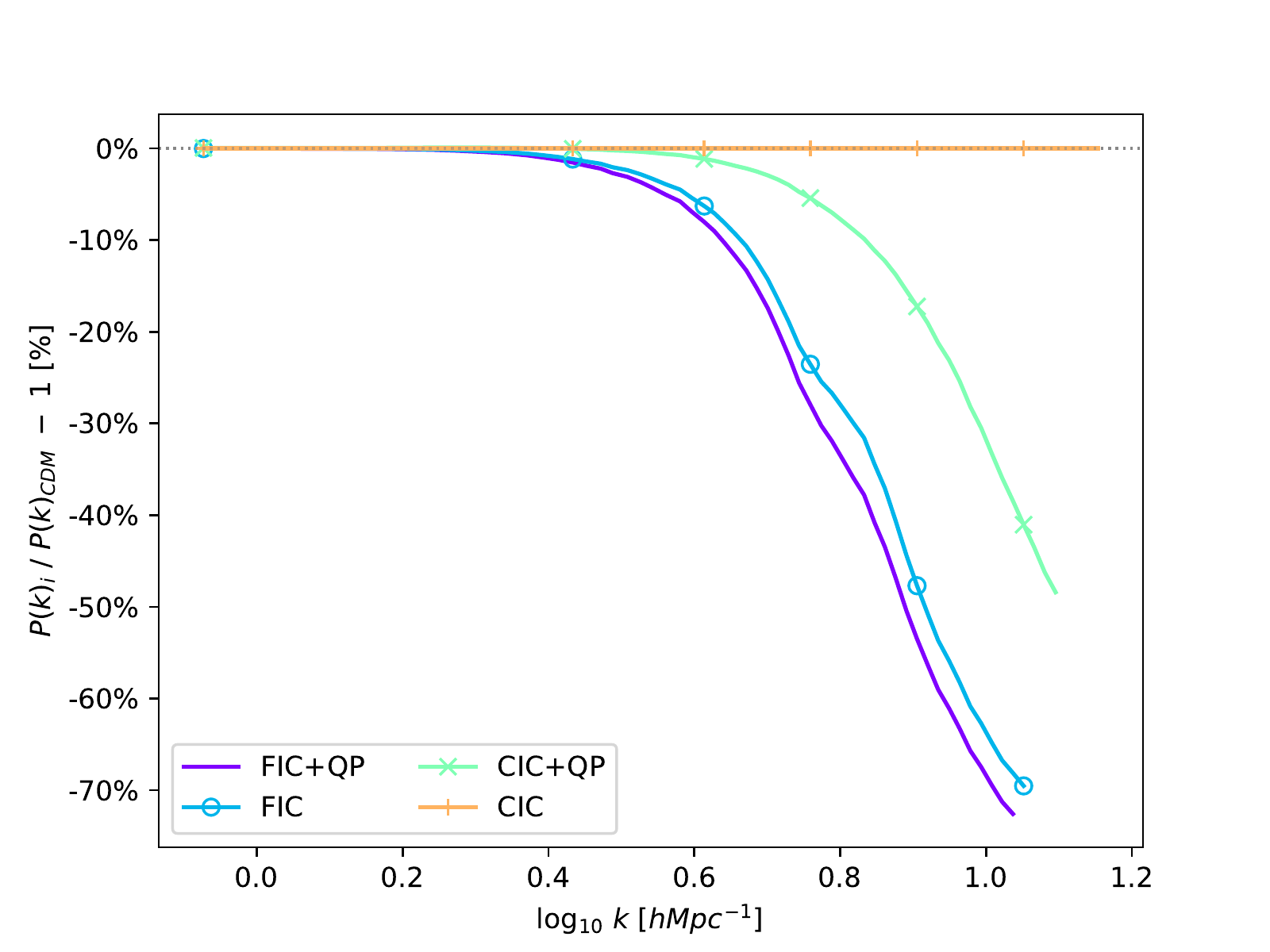} &
\includegraphics[width=\columnwidth,trim={0cm 0cm 1.5cm 1cm},clip]{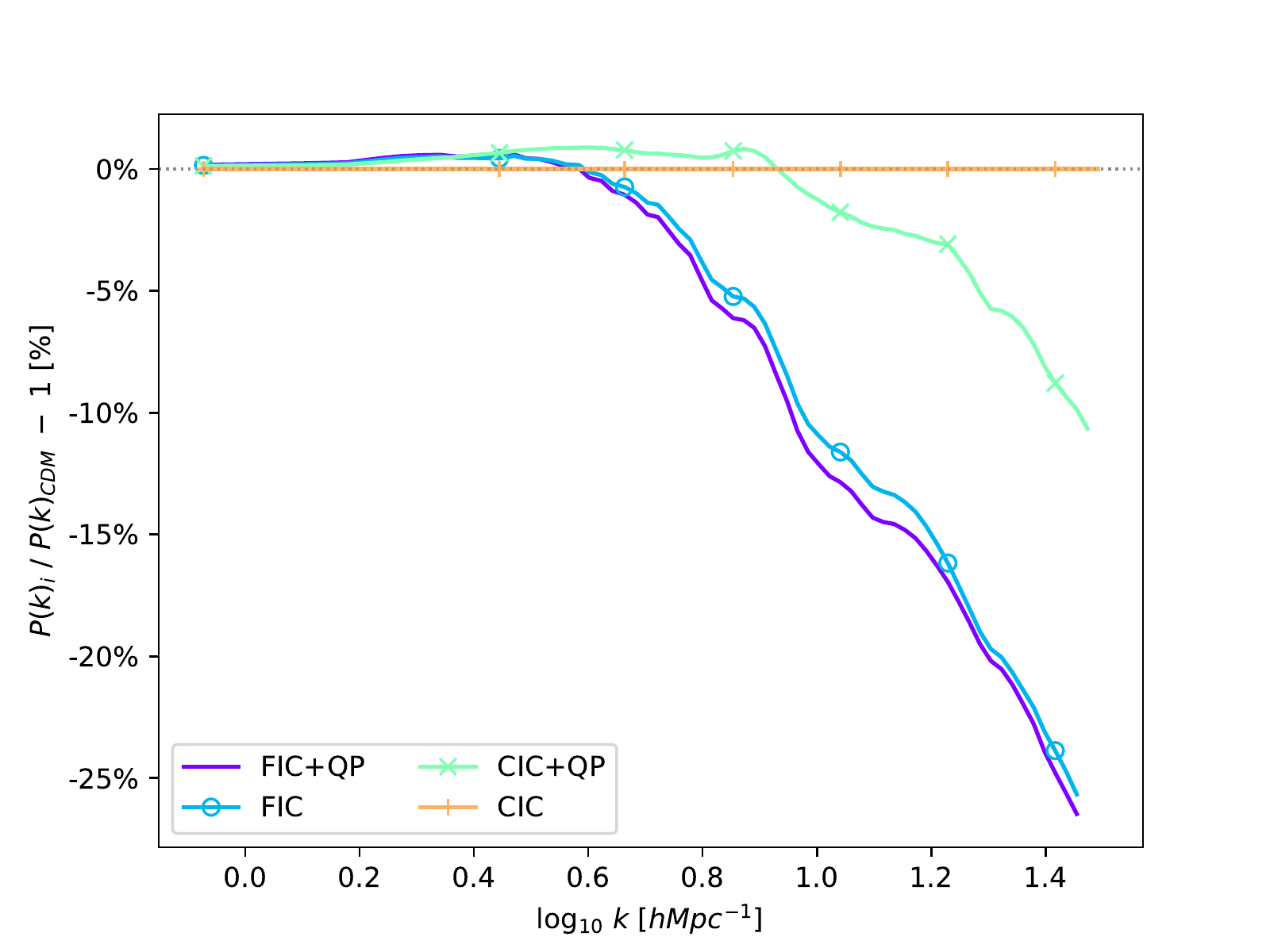} \\
\end{tabular}
\caption{Relative difference between matter power spectra between CIC, CIC$+$QP, FIC and FIC$+$QP simulations and the reference CIC setup, corresponding to standard $\Lambda $CDM.}
\label{fig:PS3}
\end{figure*}

\begin{figure*}
\begin{tabular}{cc}
\includegraphics[width=\columnwidth,trim={3.3cm 0.5cm 3.3cm 2.5cm},clip]{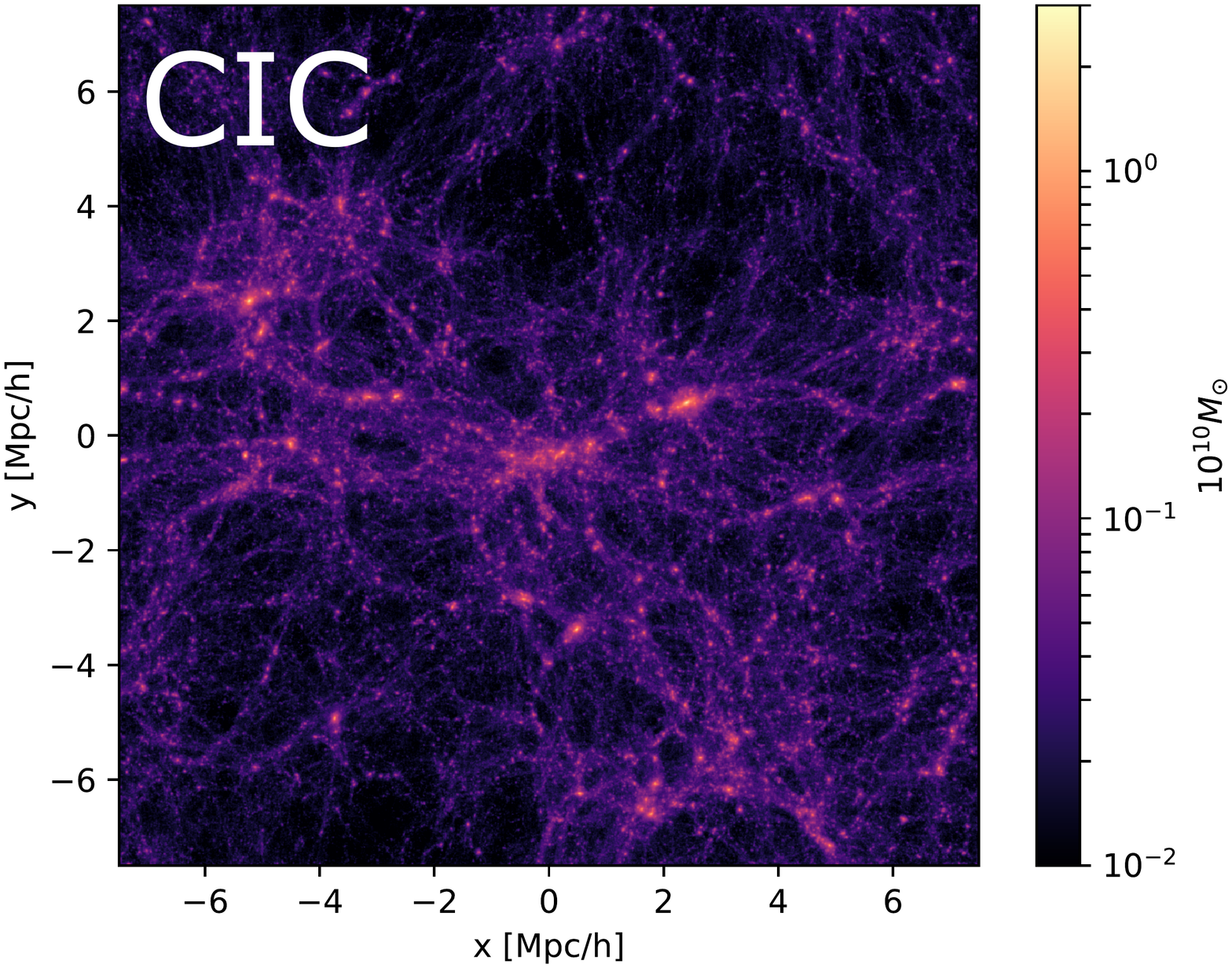} &
\includegraphics[width=\columnwidth,trim={3.3cm 0.5cm 3.3cm 2.5cm},clip]{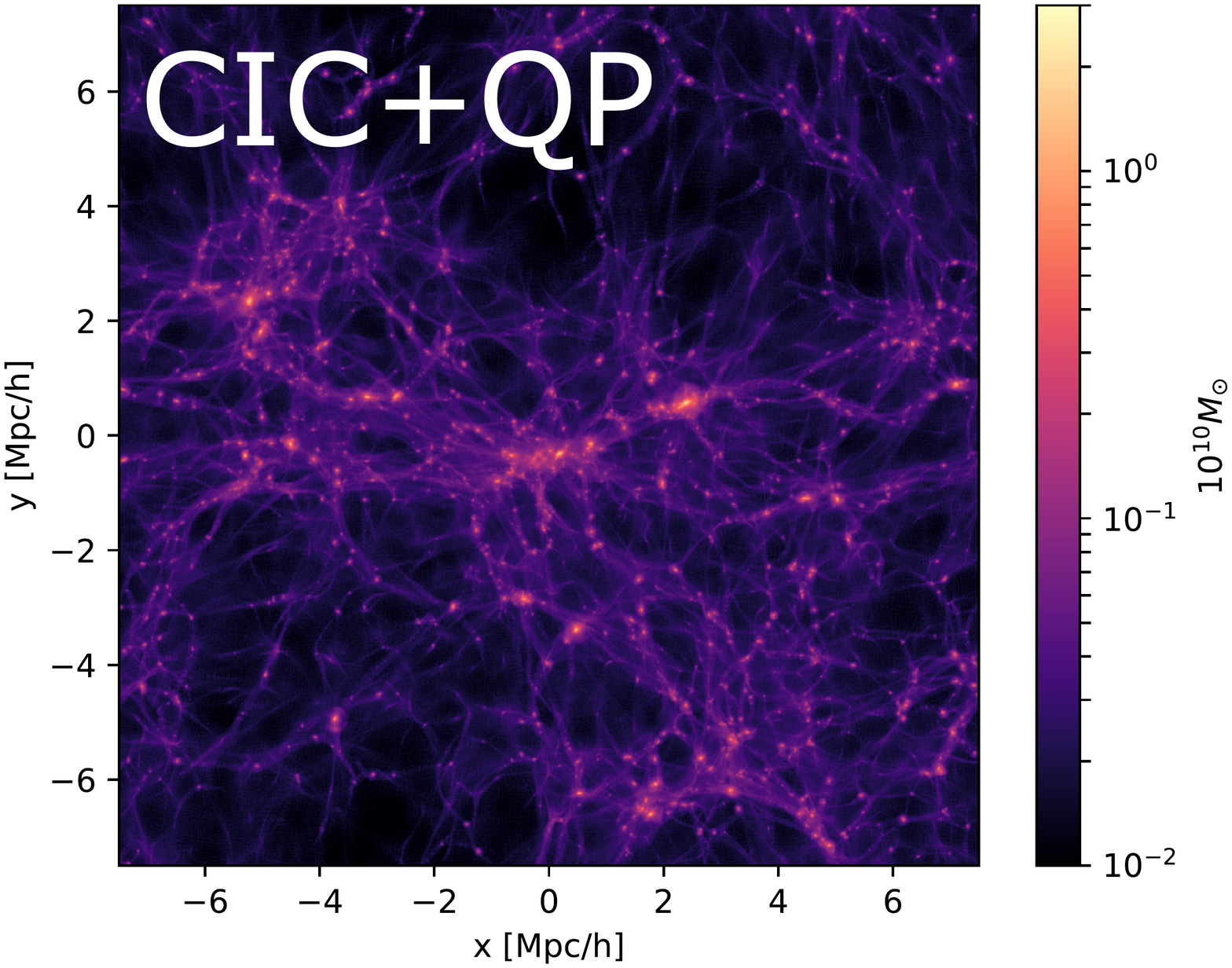} \\
\\
\includegraphics[width=\columnwidth,trim={3.3cm 0.5cm 3.3cm 2.5cm},clip]{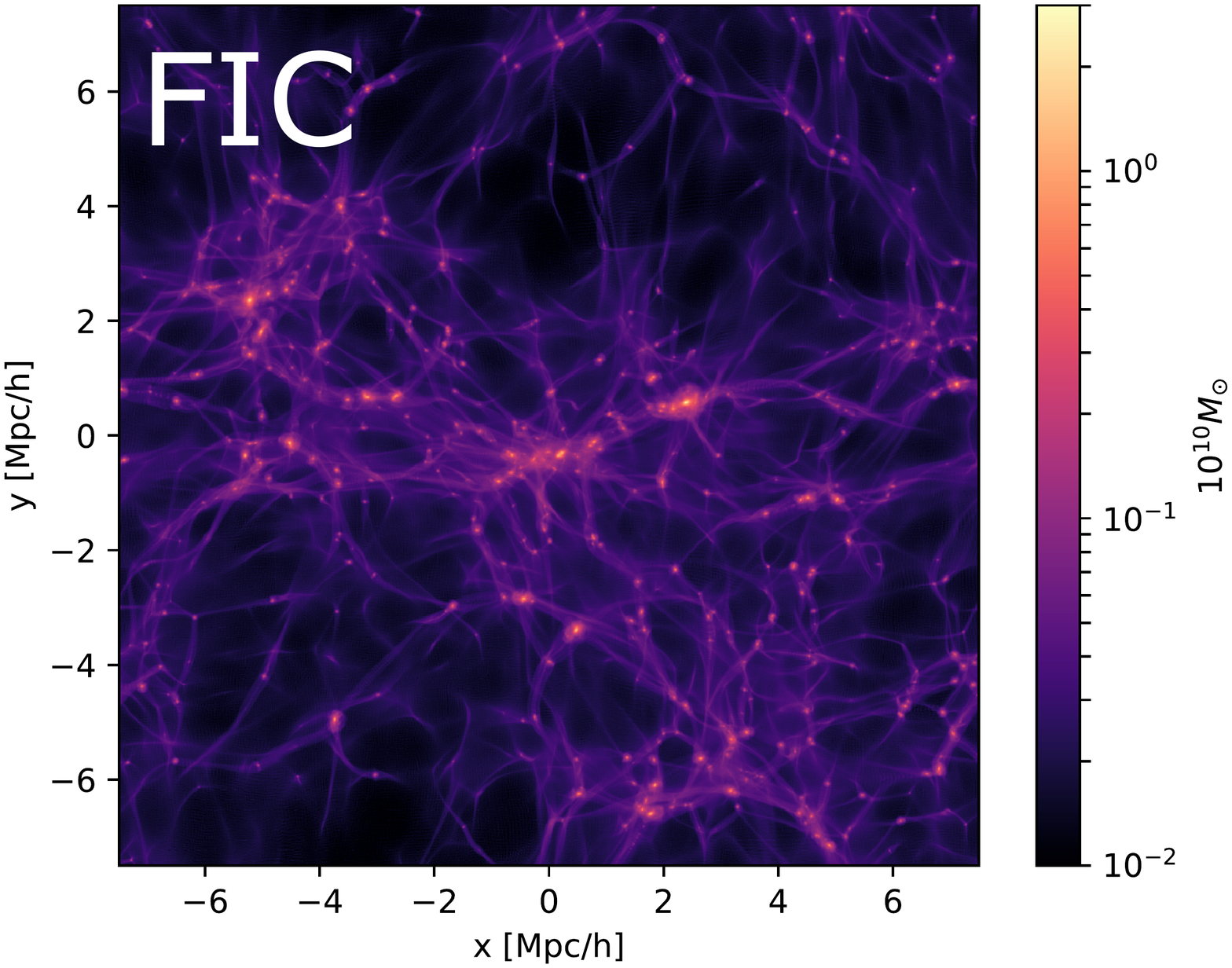} &
\includegraphics[width=\columnwidth,trim={3.3cm 0.5cm 3.3cm 2.5cm},clip]{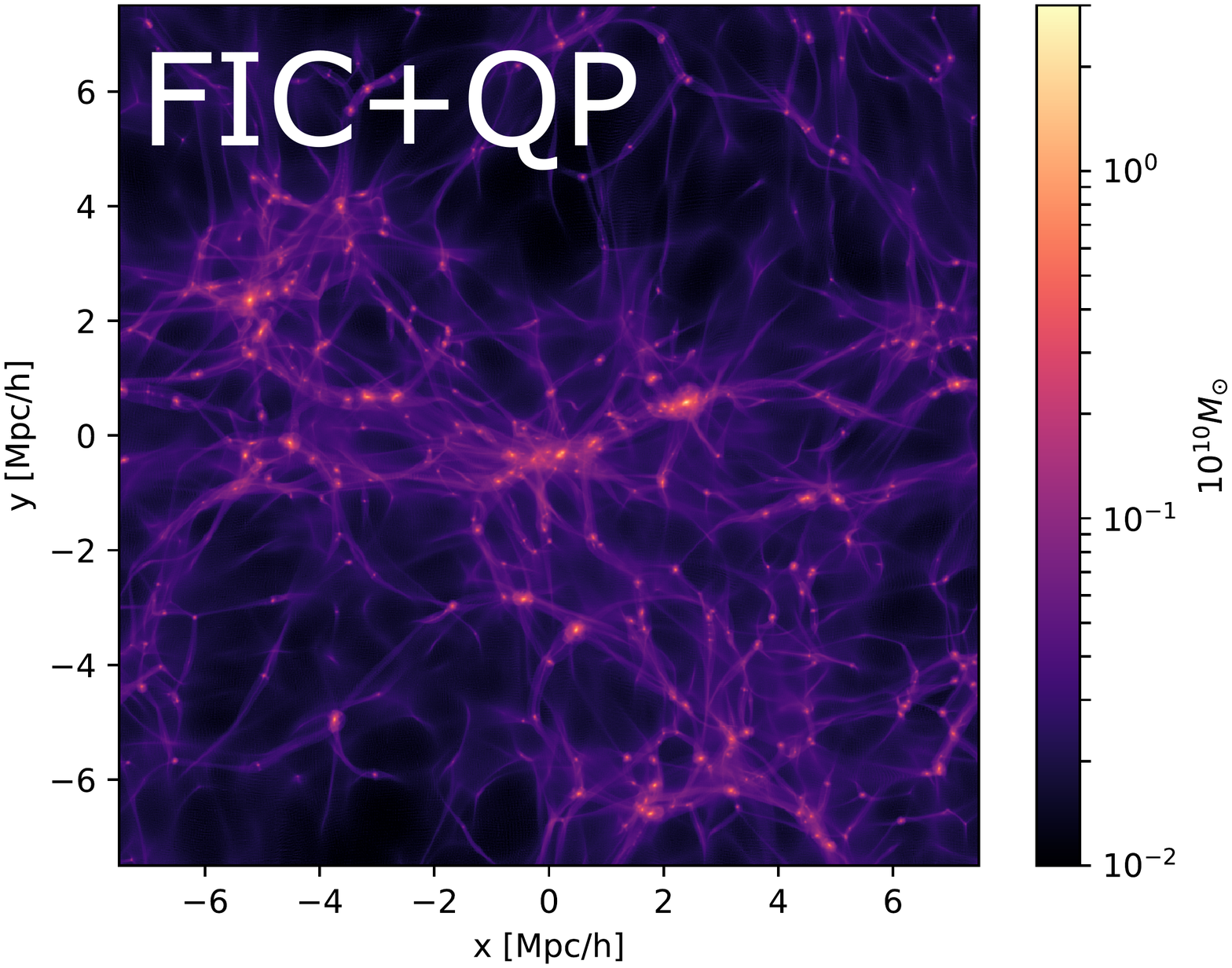} \\
\end{tabular}
\caption{Density distribution of four simulations starting from standard initial conditions (\textit{top}) or suppressed with {\small AxionCAMB} (\textit{bottom}) and evolved with (\textit{right}) or without (\textit{left}) Quantum Potential effects.}
\label{fig:MAPS3}
\end{figure*}

In Fig.~\ref{fig:PS3} we display the relative difference of the matter power spectrum in the four simulations with respect to the reference CIC run. As one can see from the plots, the four simulations are paired at the starting time in the two different initial conditions FIC and CIC ({\em top-left pane} of Fig.~\ref{fig:PS3}), and immediately decouple under the effect of the QP ({\em top-right panel}). The small-scale overdensities are either disrupted (for the CIC$+$QP case) or frozen (for the FIC$+$QP case) by the QP and this results in a drop of the power spectrum compared to the corresponding runs without QP. At lower redshifts, the maximum suppression with respect to the reference CIC run is obtained for the FIC+QP simulation, which features an additional $\approx 5-10\%$ suppression (depending on the redshifts) at small scales compared to the FIC case where the suppression is only imprinted in the initial conditions. This result seems to suggest that the approximate treatment of neglecting the QP in the dynamical evolution of a suppressed primordial matter power spectrum -- that has been widely employed in the literature \citep[see e.g.][]{Schive16,Armengaud17,Irsic17} -- may not be sufficient for precise quantitative assessments of the observational features of FDM and generalised Axion Dark Matter models. 

Since the range of action of the QP is characterised by typical lenght $k_Q(z,m)$ (see Eq.~\ref{eq:kq}), we do expect that for some range of redshifts and masses the overall linear effect of QP can indeed be accurately encoded in the initial conditions. However, our results suggest that the QP induce a suppression at redshift-dependent scales whose integrated effect cannot be overlooked, at least for scales $\lesssim 1$ Mpc$/h$. Neglecting the QP action in the dynamics may result in an overestimation of power as large as $~10\%$ at scales $\approx 300$~ kpc at $z=3$.
Therefore, we do conclude that accurate simulations consistently including the QP in the dynamical evolution of cosmic structures are necessary to place fully reliable constraints on the parameter space of FDM and Axion Dark Matter models.

Our results are in stark contrast with the previous findings of \citet{Veltmaat16}, who found that the matter power spectrum obtained from simulations with suppressed initial conditions (performed with an adaptive mesh refinement code for FDM cosmology) is enhanced at small scales by the action of the QP rather than being further suppressed.
Even if the QP can indeed be attractive in small regions of space around overdensities -- as discussed in section~\ref{sec:cv} -- we find that the overall integrated effect on the density field is always opposite to gravity -- therefore to matter gravitational collapse -- thus resulting in a suppression of small-scale power whose intensity may be amplified by using already suppressed initial conditions rather than being overturned by it.

In Fig.~\ref{fig:MAPS3} we provide a visual comparison of the large-scale matter distribution in our four test simulations by showing maps of the density field at redshift $z=3$ for the four runs. It is evident how most of the low-mass structures that appear in the CIC setup are absent in the other simulations. Both the suppression imprinted in the initial conditions and the one resulting from the QP effect alone are able to wipe out inhomogeneities and prevent dark matter from accrete on small-scale structures, the former being more effective than the latter in this regard. However, the combined effect of suppressed initial conditions and of QP acting on the dynamics of dark matter particles in the simulation FIC+QP -- which corresponds to the most realistic and self-consistent setup for FDM -- is found to provide the strongest impact on the abundance of low-mass objects.

\subsection{Performance}
\label{sec:perf}

In this Section we briefly describe the overall performance of \AG for the cosmological runs described above, and we compare it to a standard CDM simulation performed with the unmodified version of \G.

The SPH implementation that computes the QP and its contribution to particle acceleration with three cycles on all the FDM particles has been built analogously to the extremely optimized baryonic one, in order to spread the computation and memory load across the CPUs. Therefore, \AG should perform in a similar way to a hydrodynamical simulation with no CDM particles. As a consequence, the overhead compared to a collisionless CDM-only simulation is still significant, but definitely much weaker compared to grid-based FDM full-wave solvers \citep[such as e.g.][]{GAMER}.

\begin{figure*}
\begin{tabular}{cc}
\includegraphics[width=\columnwidth,trim={0.5cm 0.7cm 1.7cm 1.7cm},clip]{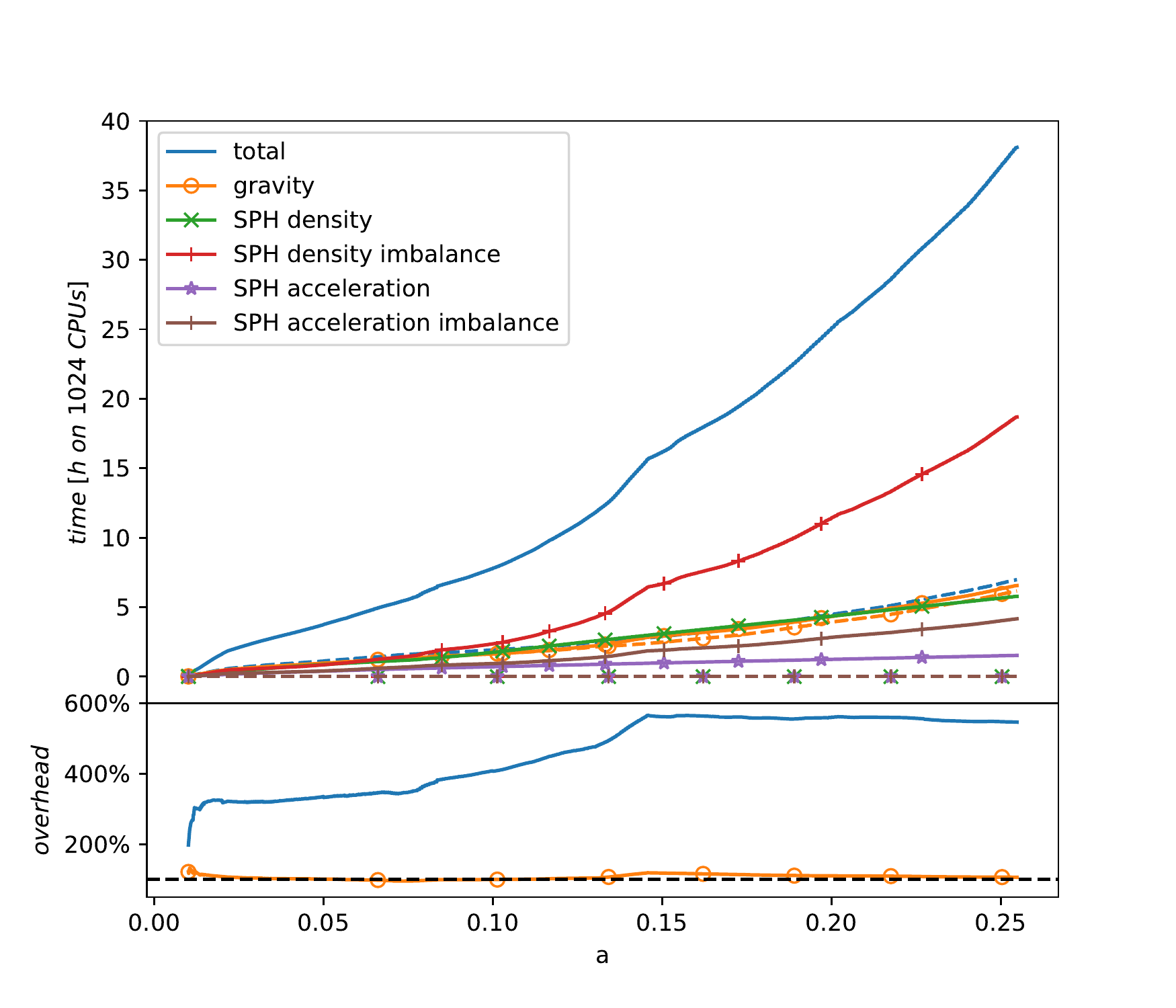}
\includegraphics[width=\columnwidth,trim={0.5cm 0.7cm 1.7cm 1.7cm},clip]{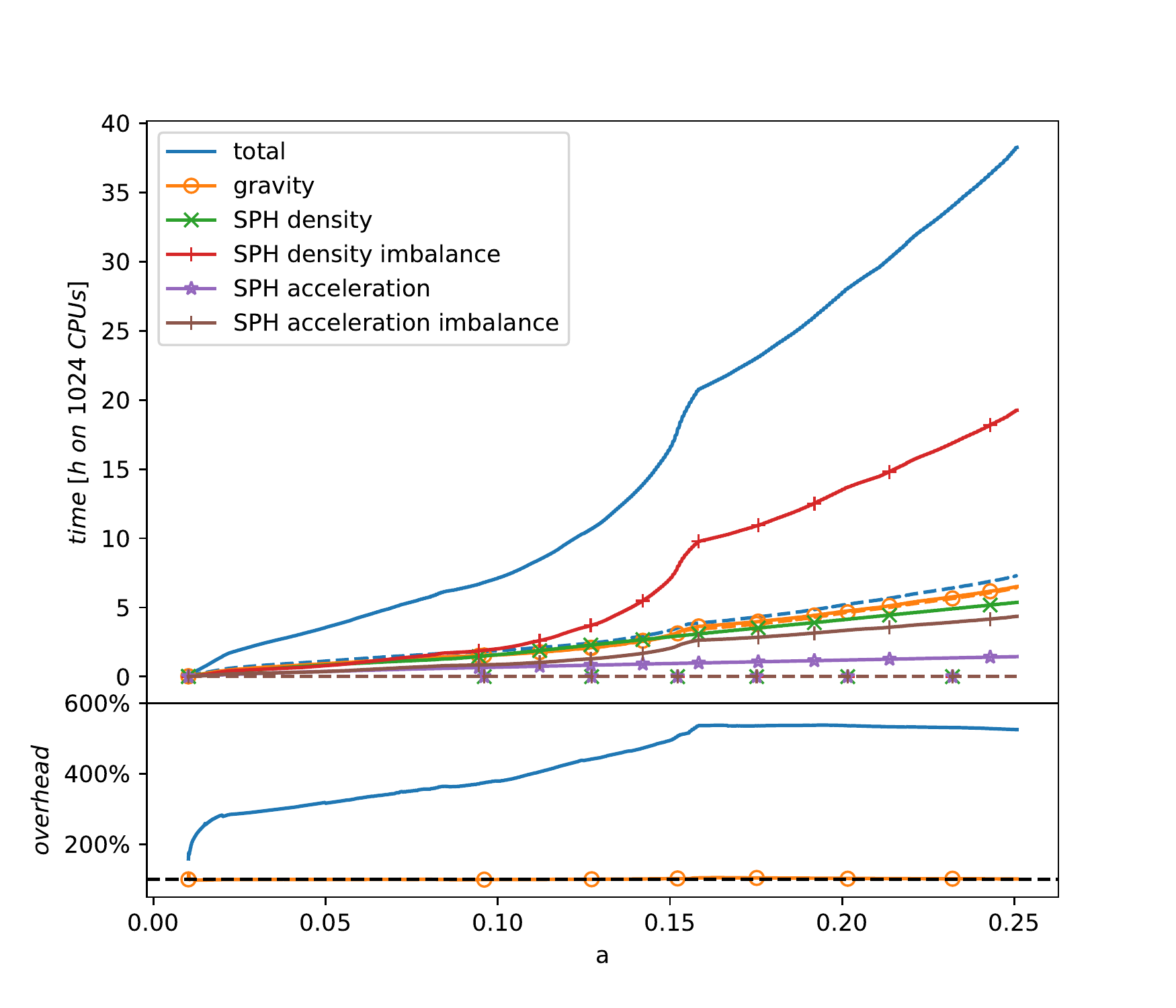}
\end{tabular}
\caption{CPU time spent as a function of the scale factor $a$ for the simulation with QP ({\em solid lines}) and without it ({\em dashed lines}) starting from CDM (\textit{left}) and FDM (\textit{right}) initial conditions. The total CPU time is plotted together with the tree-gravity and SPH routines for density and QP acceleration contributions --and relative imbalance between CPUs--. The bottom panels show the overhead for each contribution, as defined by $\text{time}_\text{solid}/\text{time}_\text{dashed}$.}
\label{fig:CPU}
\end{figure*}

In Fig.~\ref{fig:CPU} we show the CPU time and the overhead for the CIC, CIC$+$QP, FIC and FIC$+$QP simulations, paired with respect to initial conditions to highlight the additional computational load of the QP computation. Contributions of the routines of gravity solver are presented, along with SPH routines devolved to the bare computation of the density derivatives, the QP acceleration acting on particles and the respective imbalance between CPUs.

As we can see, starting from CDM initial conditions ({\em left panel}) results in an overhead of a factor of $\sim 3$ right from the beginning of the simulation in the case when the QP is included. This is due to extra work --needed to compute the QP-- arising from the reaction to the out of equilibrium configuration provided in the initial conditions. As one can see from the figure, this overhead only weakly grows during the remainder of the simulation up to a total factor of $\sim 5$ at $z=3$. When FDM initial conditions are used ({\em right panel}), the overburden is indeed less pronounced in the early phases of the evolution whereas the final computational time is a factor $\sim 5$ larger than the case without QP also in this case.

We find that the major contribution to CPU time is the one associated to the imbalance between CPUs in the SPH calculation --namely {\em SPH density imbalance} and {\em SPH acceleration imbalance}--, while the time spent for the bare SPH calculation -- {\em SPH density} and {\em SPH acceleration} -- make up for less than $20\%$ of the total time of the simulation.

Therefore, we conclude that given the relatively low overhead obtained for simulations starting from suppressed initial conditions, the inclusion of the QP in the dynamics implemented in \AG -- as would be required from theory -- does not affect dramatically the performance and the feasibility of large cosmological simulations while contributing with important physical information.

\section{Conclusions}
\label{sec:conclusions}

We have presented an extension of the massively parallel N-body code \G for non-linear simulations of Fuzzy and Axion dark matter cosmologies based on the solution of the dynamic Schr\"odinger equation in the Madelung formulation through Smoothed Particle Hydrodynamics techniques. Our code, that we called \AG, shares the same general structure of \G thereby inheriting its scalability and load-balance efficiency, as well as the wealth of additional implementations -- ranging from sophisticated algorithms for radiative gas physics to Dark Energy and Modified Gravity modules -- that have been included in \G over the past years.

More specifically, our implementation of Fuzzy Dark Matter is based on the solution of the associated Quantum Potential (see Eq.~\ref{eq:QP}) via a series of spatial derivatives of the density field computed from each simulation particle through the standard SPH kernel. Nonetheless, the higher order of spatial derivatives compared to standard SPH simulations that is required to compute the Quantum Potential acceleration (third order derivatives of the density field compared to the first order derivatives required for standard hydrodynamical forces) results in a very poor accuracy of the solver if the standard approach of \G for the computation of spatial gradients is recursively employed. To overcome this problem, we have explored alternative methods to compute higher-order derivatives based on a regularisation of each first-order derivative in regions of constant density (Eq.~\ref{eq:1SPHcorrect}). 

This improved scheme provides much more accurate and stable results for the computation of the Quantum Potential and of its associated acceleration, as we demonstrated through a series of tests for density distributions with a known Quantum Potential analytical solution.

First of all, we investigated a one-dimensional density front described by a hyperbolic tangent shape in a three-dimensional box, realised either by changing the mass of tracer particles set on a regular cartesian grid or by moving particles of constant mass to reproduce the desired density distribution. In both cases, the standard SPH approach recovers with excellent accuracy the input density, but for the latter it fails in capturing faithfully the shape of the associated Quantum Potential. On the contrary, our improved scheme provides a much more accurate solution thereby giving rise to a better representation of the resulting acceleration.

Secondly, we tested the code on a density distribution that more closely resembles the situation of a collapsing dark matter halo in cosmological simulations, namely a three-dimensional Gaussian density profile, placed at the center of a non-periodic box, again realised both by an individual mass change for a regular grid of particles and by moving around particles of equal mass. Also in this case, the standard SPH approach perfectly recovers the input density, but fails to reproduce accurately the Quantum Potential for an inhomogeneous distribution of equal mass particles. Again, our improved scheme shows much better convergence to the expected analytical solution.

Based on the success of these analytical tests, we moved to apply our algorithm to more realistic cosmological setups.
As a first test, we investigated the impact of the Quantum Potential by running two simulations with identical initial conditions, generated for a standard Cold Dark Matter power spectrum, at very high redshift ($z_{i}=999$), with and without the contribution of the Quantum Potential. This test showed a sudden re-arrangement of particles right at the start of the simulation when the Quantum Potential is included, resulting in a strong suppression of the density power spectrum at the smallest scales probed by our box compared to the standard case. This is due to having set the system -- in the case with Quantum Potential -- out of its equilibrium configuration by using a Cold Dark Matter power spectrum to generate the initial conditions. After this first phase of dramatic evolution, however, the system finds its new equilibrium configuration and starts evolving in a more relaxed way, slowly restoring small-scale perturbations during the cosmological evolution. Most importantly, the evolution is found to recover the theoretically expected linear suppression of the matter power spectrum at intermediate and low redshifts, thereby providing a positive test for the stability and the accuracy of our algorithm.

Then, we moved to compare the impact of the Quantum Potential to the effect of introducing its associated small-scale power suppression in the initial conditions, which has been claimed in the literature to be a sufficient and much cheaper approach to structure formation in Fuzzy Dark Matter cosmologies. To this end, we ran four cosmological simulations starting from a lower initial redshift $z_{i}=99$, two of which starting from standard Cold Dark Matter initial conditions, while the other two starting from a random realisation of a suppressed matter power spectrum according to the linear predictions for a given Fuzzy Dark Matter particle mass. For each of these two initial setups, we then evolved the simulations either with or without including the Quantum Potential. 

This further test showed that indeed including the Quantum Potential in the dynamics provides a qualitatively similar suppression of the matter power spectrum as one would get from just evolving linearly suppressed initial conditions. However, when both are included in the same simulation -- which represents the most consistent setup for the evolution of the system -- the resulting matter power spectrum at low redshifts shows an additional suppression of about $5-10\%$ compared to the case with no Quantum Potential. This result demonstrates that a proper implementation of the Quantum Potential in the dynamics is necessary for precision cosmology, and in particular for accurate predictions aimed to place constraints on the Fuzzy Dark Matter particle mass. Furthermore, one can expect that such additional suppression would result in more pronounced effects at the level of the structural properties of dark matter halos, which we did not investigate in this work but that will be discussed in a forthcoming paper.

Finally, we have shown that the overall performance of \AG does not make high-resolution cosmological simulations prohibitive, with an overhead compared to standard collisionless simulations of a factor of $5-6$, thereby having a moderate increase of the computational cost compared to standard SPH simulations.
\ \\

To conclude, we have presented the \AG code featuring an efficient and accurate implementation of the Quantum Potential that characterises Fuzzy Dark Matter models and in particular Ultra Light Axion particles as candidates for the cosmological budget of dark matter. We have described the algorithm implemented in the code, the strategies we adopted to improve its accuracy compared to standard SPH techniques, and shown tests for analytical density distributions. We have also employed the code for realistic cosmological simulations, showing that a consistent treatment of the Quantum Potential in the dynamical evolution of the system is necessary to account for the full suppression of power at small scales that represents the most prominent observational feature of Fuzzy Dark Matter scenarios.

\section*{Acknowledgements}

We acknowledge support from the Italian Ministry for Education, University and Research (MIUR)
through the SIR individual grant SIMCODE, project number RBSI14P4IH. The simulations described in this work have been performed on the Marconi supercomputer at Cineca thanks to the PRACE allocation 2016153604.
We are deeply thankful to Volker Springel for his valuable comments on the draft and for hosting MN at HITS during part of the development of this work. We are also thankful to David Marsh, Jens Niemeyer, Jan Veltmaat, Lam Hui and  Matteo Viel for inspiring discussions on our implementation.




\bibliographystyle{mnras}
\bibliography{BIB,baldi_bibliography}



\appendix

\section{Varying smoothing length}
\label{sec:dhsml}

In order to preserve energy and entropy conservation of the algorithms -- at least in the appropriate limits -- it is imperative to take into account the terms arising from the variation of the smoothing length $h$ required to satisfy Eq.~\ref{eq:NN}. 
To do so, we follow the approach described in \citet[][]{Springel01}, where Lagrangian multipliers are introduced to keep track of $h$-derivative terms.

Let us consider a generic Lagrangian for a FDM N-body ensemble with the form
\begin{equation}
\begin{split}
\Lagr(\vec{q}, \dot{\vec{q}})=\ \sum^N_{i=0} \ \frac 1 2 & m_i |\dot{\vec{r_i}}|^2- m_i \frac {P_i} {{\rho_i}^2} \\
+&\frac {\hbar^2}{2m_\chi^2} m_i \frac {\nabla^2 \sqrt{\rho_i}}{\sqrt{\rho_i}}+\lambda_i (V_i\rho_i - M)
\end{split}
\end{equation}
expressed in terms of the variables $\vec q_i = (\vec r_i, h_i)$ and where the different terms represent the kinetic energy, the self-interaction between particles -- described through a pressure function $P$ -- and the QP contribution. The last term enforces Eq.~\ref{eq:NN} through N Lagrangian multipliers $\lambda$.

The set of equations of motion linked to the multipliers, one for every $j$ particle in the ensemble, results in
\begin{equation}
\lambda_j=(1- \frac 1 {f_j}) \frac {m_j} {V_j} \left[\frac {P_j}{\rho_j^2} - \frac{\hbar^2}{m^2_\chi} \de{\rho_j} \left( \frac {\nabla^2 \sqrt{\rho_j}}{\sqrt{\rho_j}} \right) \right]
\end{equation}
where we defined the parameters $f$ as
\begin{equation}
f_j = \left( 1+ \frac h {3\rho_j} \de{h}\rho_j \right)
\end{equation}
that we use in the text \citep[notice that our definition of $f$ in the inverse with respect to the one in][]{Springel01}.
Substituting the Lagrangian multipliers, the set of equation of motion related to the positions $\vec r$ can be expressed as
\begin{equation}
m_i \ddot{\vec{r_i}}=- \sum^N_{j=0} \frac {m_j} {f_j} \frac {P_j} {\rho_j^2} \vec \nabla \rho_j
+\frac {\hbar^2}{2m_\chi^2} \frac {m_j} {f_j} \vec \nabla \left( \frac {\nabla^2 \sqrt{\rho_j}}{\sqrt{\rho_j}} \right)
\end{equation}
which then can be implemented through SPH algorithms.
To summarize, the adaptive adjustment of the smoothing lengths of each single particle contributes to particle accelerations through terms involving $h$-derivatives that can be expressed as $f$ factors in the SPH neighbours summation. 

\section{Quantum Potential in the case of a Gaussian kernel}
\label{sec:zhang}

\citet[][]{Zhang16} have recently presented a technique to approximate the particle-particle interaction induced by the QP in N-Body simulations. Even if their technique to describe fluid interactions is not explicitly derived from SPH, they share the same idea of approximating the real density field at particle positions with discrete sums over a small particle ensemble enclosed in an effective volume.

The authors start from a continuous FDM Lagrangian in the non-interacting case, reading
\begin{equation}
\Lagr(q,\dot q)=\frac 1 2 \int \rho |\dot {\vec r}|^2 dx^3 - \frac {\hbar^2}{2m_\chi^2} \int |\vec \nabla \sqrt{\rho}|^2 dx^3
\end{equation}
where we recognize the kinetic and the QP terms.

To discretize the equation above, they approximate the integrals with sums on fluid particles $j$ which are mathematically represented with a Gaussian kernel.
The volumes $V_j$, therefore, represent the typical volume occupied by each particle linked to the mass and the smoothing length $h_j$ of the kernel used.
The two terms in the Lagrangian in the discretized form read
\begin{equation}
\label{eq:discretization}
\begin{split}
\int \rho |\vec v|^2 dx^3 \rightarrow& \ \sum_j \rho_j |\vec v_j|^2 V_j \\
\int |\vec \nabla \sqrt{\rho}|^2 dx^3 \rightarrow& \ \sum_j |\vec \nabla \sqrt{\rho_j}|^2 V_j
\end{split}
\end{equation}

In \citet[][]{Zhang16}, while the discretization of Eq.~\ref{eq:discretization} is carried out correctly, the same cannot be said for the equation of motion derived from them. As a matter of fact, the particle volumes $V_j$ -- that by definition correspond to the ratio between the mass and the density $m_j/\rho_j$ -- are not taken into account correctly in the derivation of the QP discretized term, effectively setting that $\vec \nabla V_j = 0$. The authors choose to treat the particle volumes $V_j$ as constants, thus effectively choosing a single smoothing length for all the particles, introducing a correction factor $\mathit{B}_j$ to account for the specific volume occupied by the particle. The $\mathit{B}_j$ factors are fitted {\it{a posteriori}}, ending up being proportional to the smoothing length cube, as one would expect from Eq.~\ref{eq:NN}. However, the authors neglect the volume derivatives arising from $\mathit{B}_j$ which translates in the lack of the density term in the denominator in the very definition of Eq.~\ref{eq:QP}.

Using Gaussian kernel functions with smoothing lenghts $h$ such that
\begin{equation}
\rho_i \propto \sumnn \frac{m_j}{(2\pi h_j^2)^{3/2}} e^{-r_{ij}^2 / h_j^2}
\end{equation}
the QP contribution to acceleration between two particles $i$ and $j$ at a distance $r$ with this wrong approach is
\begin{equation}
\ddot {\vec r}\ |_{QP} = \frac {\hbar^2}{2m_{\chi}^2} \frac{V_j}{m_i} \vec \nabla \left( |\vec \nabla\sqrt{\rho_j}|^2 \right) = \frac {\hbar^2}{2m_{\chi}^2} \frac{m_j}{m_i} \frac {\vec r }{h_j^4} \left( 1- 2\frac {r^2 }{h_j^2} \right)
\end{equation}
whereas the right one including the volume derivative is
\begin{equation}
\ddot {\vec r}\ |_{QP}  = \frac {\hbar^2}{2m_{\chi}^2} \frac{1}{m_i} \vec \nabla \left(|\vec \nabla\sqrt{\rho_j}|^2 \ V_j \right) = \frac {\hbar^2}{2m_{\chi}^2} \frac{m_j}{m_i} \frac {\vec r }{h_j^4} 
\end{equation}

The different regimes originated by the change in sign of the quantum interaction are, therefore, a fictitious effect given by an error in the discretization scheme.

The counter-intuitive attractive behaviour at small distance found by the authors does not match the physical description of the process, as we know that the QP would repel two free point-like particles -- recovered from the equations above in the limit of $h \rightarrow 0$ -- irrespectively from their separation and thus always opposite to gravity.



\bsp	
\label{lastpage}
\end{document}